%% file: paper.tex
\documentclass[sigconf,10pt,screen,acmthm]{acmart}

\input{preamble}

\input{macros}

\acmConference[]{}{}{}
\newtoggle{arXiv}
\toggletrue{arXiv}
\newtheorem*{prop*}{Theorem}

\begin{document}
\title{An Elementary Introduction to Kalman Filtering}

\author{Yan Pei}
\affiliation{University of Texas at Austin}
\email{ypei@cs.utexas.edu}

\author{Swarnendu Biswas}
\affiliation{Indian Institute of Technology Kanpur}
\email{swarnendu@cse.iitk.ac.in}

\author{Donald S. Fussell}
\affiliation{University of Texas at Austin}
\email{fussell@cs.utexas.edu}

\author{Keshav Pingali}
\affiliation{University of Texas at Austin}
\email{pingali@cs.utexas.edu}

\settopmatter{printacmref=false, printccs=true, printfolios=false}
\setcopyright{none}
\pagestyle{plain} % removes running headers
\renewcommand\footnotetextcopyrightpermission[1]{} % removes footnote with conference information in first column

\begin{abstract}
	% \francis{\\ Highlight text \textbackslash hl\{text\} \\
	% 	Highlight equations \textbackslash mathcolorbox\{color (e.g. yellow)\}\{equation\} \\
	% 	Remove some text \textbackslash st\{text\}
	% }

	%CACM submission todo list:
	%\begin{itemize}
	%	\item \st{Section 1\&2: change real estate example to temperature sensor example with and without systematic error (Swarnendu)}
	%	\item \st{Remove some of "statistical" (Francis)} (Some left)
	%	\item \st{Add expectation, covariance definition to appendix (Francis)}
	%	\item \st{Use term ``Innovation'' (Keshav)}
	%	\item \st{Incorporate 6.3 to 6.2.1 using a different example (Francis)}
	%	\item \st{EKF and UKF brief introduction (Francis)}
    %    \item \st{Use small-hat}
    %    \item \st{Revise EKF and UKF}
    %    \item \st{Optimality of Equation 34. Proof solved.}
    %    \item \st{Systematic error in BLUE and in data fusion from reviewer 1}
	%\end{itemize}

	%On the stack for CACM submission:
	%\begin{itemize}
	%	\item Solve review 2.3 after Josep's response
	%	\item Shrink the length of the equations
	%\end{itemize}

	%On the stack for tech-report
	%\begin{itemize}
	%	\item Add one section to derive kalman filter using scalar
	%	\item Explain whether system model noise is uncorrelated with a priori estimate from last time step (Proof is available, but we need intuitive explanation)
	%\end{itemize}

	% \swarnendu{We have agreed to make two versions of the paper: one for CACM and one as an extended technical report published on arXiv and referred from the CACM paper. We will first prioritize the CACM submission. Also, we can have different builds by using toggle macros.}

Kalman filtering is a classic state estimation technique used in
application areas such as signal processing and autonomous control of vehicles. It
is now being used to solve problems in computer systems such as controlling
the voltage and frequency of processors.

Although there are many presentations of Kalman filtering in the literature,
they usually deal with particular systems like autonomous robots or linear systems
with Gaussian noise, which makes it difficult to understand the general principles
behind Kalman filtering. In this paper, we first present the abstract ideas behind
Kalman filtering at a level accessible to anyone with a basic knowledge of
probability theory and calculus, and then show how these concepts can be
applied to the particular problem of state estimation in linear systems.
This separation of concepts from applications should make it easier to understand Kalman filtering and to apply it to other problems in computer systems.
\end{abstract}



\maketitle

\section{Introduction}
\label{sec:intro}
\input{intro}

\section{Formalization of estimates}
\label{sec:formal}
\input{formal}

\section{Fusing Scalar Estimates}
\label{sec:scalar}
\input{scalar}

\section{Fusing Vector Estimates}
\label{sec:vector}
\input{vector}

\section{Best linear unbiased estimator (BLUE)}
\label{sec:hidden}
\input{hidden}

\section{Kalman filters for linear systems}
\label{sec:dynamics}
\input{dynamics}
\section{Extension to nonlinear systems}
\label{sec:extension}
\input{extension}

\section{Conclusion}
\label{sec:conclusion}
\input{conclusions}

\begin{acks}
	This research was supported by NSF grants 1337281, 1406355, and 1618425, and by DARPA contracts FA8750-16-2-0004 and FA8650-15-C-7563. The authors would like to thank
K. Mani Chandy (Caltech), Ivo Babuska (UT Austin) and Augusto Ferrante (Padova) for their feedback on this paper.
\end{acks}

\bibliographystyle{ACM-Reference-Format}
\bibliography{paper}

\newpage
\appendix
\appendixpage
\begin{appendix}
\label{sec:appendix}
	\input{appendix}

\end{appendix}

%\section{Appendix: Matrix Derivatives}
%\label{sec:appendix}
%\input{appendix}

\end{document}

%% file: preamble.tex
\usepackage{etoolbox}
\usepackage{booktabs} % For formal tables
\usepackage{algpseudocode}
\usepackage{algorithm}
\usepackage{listings}
\usepackage{mathtools}
\usepackage{amsmath}
\usepackage{amssymb}
\usepackage{bm}
\usepackage{enumerate}
\usepackage[framemethod=TikZ]{mdframed}
\usepackage{xspace}
\usepackage{subfig}
\usepackage{graphicx}
\usepackage{pgfplots}
\usepackage{pgfplots}
\usepackage{dcolumn}
\usepackage{widetext}
\usepackage{soul}
\pgfplotsset{compat=1.10}
\usepgfplotslibrary{units} % Allows to enter the units nicely
\usepackage{cleveref}
\mdfdefinestyle{MyFrame}{%
	linecolor=blue,
	outerlinewidth=2pt,
	roundcorner=20pt,
	innertopmargin=\baselineskip,
	innerbottommargin=\baselineskip,
	innerrightmargin=20pt,
	innerleftmargin=20pt,
	backgroundcolor=gray!50!white}
\usepackage{siunitx}
\usepackage{textcomp}
\usepackage[toc,page]{appendix}

\usepackage{comment}
\usepackage{relsize}
\usepackage[utf8x]{inputenc} % utf8 encoding
\usepackage[T1]{fontenc} % use T1 fonts
\usepackage{bm} % bold math
\usepackage{color} % change text color
\usepackage{tikz}
\usetikzlibrary{arrows,shapes,decorations,automata,backgrounds,calc,topaths, positioning}
\usepackage{verbatim}
\hypersetup{
 hidelinks,
 colorlinks=false
}

%% file: macros.tex
\newenvironment{closeenumeratei}
{\begin{enumerate}[(i)]
    \setlength{\itemsep}{1pt}
    \setlength{\parskip}{0pt}
    \setlength{\parsep}{0pt}}
{\end{enumerate}}

\lstdefinestyle{python}{
  language=Python,
  basicstyle=\ttm,
  otherkeywords={self},             % Add keywords here
  keywordstyle=\ttb\color{deepblue},
  emph={MyClass,__init__},          % Custom highlighting
  emphstyle=\ttb\color{deepred},    % Custom highlighting style
  stringstyle=\color{deepgreen},
  frame=tb,                         % Any extra options here
  showstringspaces=false            %
}

\lstdefinestyle{bash}{
  language=bash,
  basicstyle=\small\sffamily,
  frame=tb,
  columns=fullflexible,
  backgroundcolor=\color{blue!5},
  linewidth=1.0\linewidth,
  xleftmargin=0.1\linewidth,
  numbers=none
}

\lstdefinestyle{java}{
  language=Java,
  basicstyle=\footnotesize\sffamily,
  numberstyle=\scriptsize,
  numbersep=5pt,
  tabsize=2,
  extendedchars=true,
  breaklines=true,
  commentstyle=\color{darkgreen}\textit,
  keywordstyle=\color{blue}\textbf,
  escapeinside={\%*}{*)},            % if you want to add LaTeX within your code
  columns=flexible
}

\newcommand{\kvec}[1]{\textbf{\large{#1}}}
\newcommand{\kmat}{}
\newcommand{\kvecsub}[1]{\textbf{#1}}
\newcommand{\mse}{\textit{MSE}\xspace}

\newcommand{\kf}{Kalman filter\xspace}

\newcommand{\inform}{\nu}
\newcommand{\Inform}{N}

\newcommand\est[1]{\widehat{#1}}

%% file: intro.tex
Kalman filtering is a state estimation technique invented in 1960 by Rudolf E. K\'{a}lm\'{a}n~\cite{kalman-filter}. Because of its ability to extract useful information from noisy data and its small computational and memory requirements,
it is used in many application areas including spacecraft navigation, motion planning in robotics, signal processing, and wireless sensor networks~\cite{sensor-network-suvey,modeling-human-gait,Thrun:2005,kalman-techreport, hscc-2010}. Recent work has used Kalman filtering in controllers for computer systems~\cite{bergman2009,pothukuchi2016using, bard, poet-2015}.

Although many introductions to Kalman filtering are available in the literature~\cite{bayesian-estimation-kf, kalman-techreport,kalman-book-87,kalman-primer-book, kalman-filtering-matlab,ensemble-kalman,faragher2012,kalman-real-time, picci, information-fusion,kalman-analysis, alexbecker, kf-undergraduates, kf-pictures}, they are usually focused on particular applications like robot motion or state estimation in linear systems. This can make it difficult to see how to apply Kalman filtering to other problems. Other presentations derive Kalman filtering as an application of Bayesian inference assuming that noise is Gaussian. This leads to the common misconception that Kalman filtering can be applied only if noise is Gaussian~\cite{Julier04}.
\iftoggle{arXiv}{
The goal of this paper
}{
The goal of this paper\footnote{An extended version of this paper that includes additional background material and proofs is available~\cite{kalman-filter-arxiv}.}
}
is to present the abstract concepts behind Kalman filtering in a way that is accessible to most computer scientists while clarifying the key assumptions, and then show how the problem of state estimation in linear systems can be solved as an application of these general concepts.

Abstractly, Kalman filtering can be seen as a particular approach to combining approximations
of an unknown value to produce a better approximation. Suppose we use two
devices of different designs to measure the temperature of a CPU core.
Because devices are usually noisy, the measurements are
likely to differ from the actual temperature of the core. Since the devices are
of different designs, let us assume that noise affects the two devices in
unrelated ways (this is formalized using the notion of correlation in
Section~\ref{sec:formal}). Therefore, the measurements $x_1$ and $x_2$ are likely
to be different from each other and from the actual core temperature $x_c$. A
natural question is the following: is there a way to combine the information in
the noisy measurements $x_1$ and $x_2$ to obtain a good approximation
of the actual temperature $x_c$?

One {\em ad hoc} solution is to use the formula \mbox{$0.5{*}x_1 {+} 0.5{*}x_2$} to take the average of the two measurements, giving them equal weight. Formulas of this sort are called {\em linear estimators} because they use a weighted sum to fuse values; for our temperature problem, their general form is \mbox{$\beta{*}x_1 {+} \alpha{*}x_2$}.
In this paper, we use the term {\em estimate} to refer to both a noisy measurement and to
a value computed by an estimator, since both are approximations of unknown
values of interest.
%If the estimates are equal, fusing them should produce the same value, which can be ensured by requiring that $\alpha{+}\beta {=} 1$. Therefore the linear estimators of interest are convex combinations of the form $(1{-}\alpha)x_1{+}\alpha x_2$ where $0 {\leq} \alpha {\leq} 1$.

Suppose we have additional information about the two devices; say the second one uses more advanced temperature sensing. Since we would have more confidence in the second measurement, it seems reasonable that we should discard the first one, which is equivalent to using the linear estimator \mbox{$0.0{*}x_1 + 1.0{*}x_2$}. Kalman filtering tells us that in general, this intuitively reasonable linear estimator is not ``optimal''; paradoxically, there is useful information even in the measurement from the lower-quality device, and the optimal estimator is one in which the weight given to each measurement is proportional to the confidence we have in the device producing that measurement. Only if we have no confidence whatever in the first device should we discard its measurement.

Section~\ref{sec:formal} describes how these intuitive ideas can be quantified. Estimates are modeled as random samples from \emph{distributions}, and confidence in estimates is quantified in terms of the \emph{variances} and \emph{covariances} of these distributions.\footnote{Basic concepts including probability density function, mean, expectation, variance and covariance are introduced in Appendix~{\ref{sec:basic_terms}}.}
Sections~\ref{sec:scalar}-\ref{sec:hidden} develop the two key ideas behind Kalman filtering.

\begin{enumerate}
\item How should estimates be fused optimally?

Section~\ref{sec:scalar} shows how to fuse \emph{scalar} estimates such as temperatures
optimally. It is also shown that the problem of fusing more than two estimates can be reduced to the problem of fusing two estimates at a time without any loss in the quality of the final estimate.

% \smallskip\\

Section~\ref{sec:vector} extends these results to estimates that are \emph{vectors}, such as state vectors representing the estimated position and velocity of a robot.
%The basic ideas remain the same as in the scalar case except that variances of scalar estimates are replaced with covariance matrices of vector estimates.

\item In some applications, estimates are vectors but only a part of the vector can be measured directly. For example, the state of a spacecraft may be represented by its position and velocity, but only its position may be observable. In such situations, how do we obtain a complete estimate from a partial estimate?

Section~\ref{sec:hidden} shows how the \emph{Best Linear Unbiased Estimator (BLUE)} can be used for this. Intuitively, it is assumed that there is a linear relationship between the observable and hidden parts of the state vector, and this relationship is used to compute an estimate for the hidden part of the state, given an estimate for the observable part.
\end{enumerate}

Section~\ref{sec:dynamics} uses these ideas to solve the state estimation problems for linear systems, which is the usual context for presenting Kalman filters. Section~\ref{sec:extension} briefly discusses extensions of Kalman filtering for nonlinear systems.

%\hl{Basic probability theory and statistics terms used in this paper are introduced in Appendix~{\ref{sec:appendix}}.}

%%% Local Variables:
%%% mode: latex
%%% TeX-master: "paper"
%%% End:

%% file: formal.tex
This section makes precise the notions of \emph{estimates} and \emph{confidence} in estimates.

\subsection{Scalar estimates}

\pgfmathdeclarefunction{gauss2}{2}{%
  \pgfmathparse{1/(#2*sqrt(2*pi))*exp(-((x-#1)^2)/(2*#2^2))}%
}
\pgfmathdeclarefunction{gauss3}{3}{%
  \pgfmathparse{1/(#3*sqrt(2*pi))*exp(-((#1-#2)^2)/(2*#3^2))}%
}

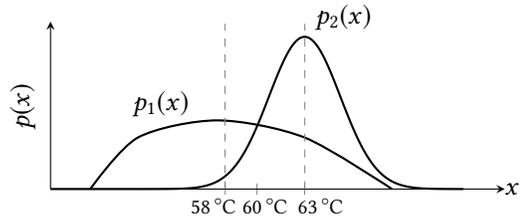
\begin{figure}[t]
  \centering
  %\subfloat[Devices with systematic and random errors]{
  \begin{tikzpicture}[scale=1]
    \begin{axis}[
    % Define probability distribution functions
    declare function={
        binom(\n,\p) = \n!/(x!*(\n-x)!)*\p^x*(1-\p)^(\n-x);
        normal(\m,\s) = 1/(\s*sqrt(2*pi))*exp(-((x-\m)^2)/(2*\s^2));
    },
      every axis plot post/.append style={smooth}, % All plots: smooth, no marks
      axis x line=bottom, % no box around the plot, only x and y axis
      axis y line=left, % the * suppresses the arrow tips
      % every axis y label/.style={at=(current axis.above origin),anchor=south},
      % every axis x label/.style={at=(current axis.right of origin),anchor=west},
      enlargelimits=upper, % extend the axes a bit to the right and top
      no markers,
      xtick=\empty,
      ytick=\empty,
      samples=100,
      clip=false,
      domain=-28:24,
      height=1.5in,
      width=3in,
      ylabel={$\mathlarger p(x)$},
      xtick={-6, -2, 4},
      xticklabels={,,},
      xtick pos=left,
    ]
    %\addplot[thick, black] {gauss2(-2,8)};

    \addplot[thick, black,
             patch, mesh,
             patch type=quadratic spline]
    coordinates {
        (-23, 0) (-17, 0.75*0.04) (-7, 0.04) %left 3 dots
        (4, 0.75*0.04) (15, 0.00) %right 2 dots
    };

    \addplot[thick, black] {gauss2(4,4.5)};

    \node[right] at (axis cs:28, 0)  {\textsl{x}};
    % \node[below] at (axis cs:-4, 0) {\footnotesize \$500K};
    % \node[below] at (axis cs:10, 0) {\footnotesize \$1 million};
    \node[below] at (axis cs:-7.5, 0) {\footnotesize \SI{58}{\celsius}};
    \node[below] at (axis cs:-1, 0) {\footnotesize \SI{60}{\celsius}};
    \node[below] at (axis cs:6, 0) {\footnotesize \SI{63}{\celsius}};
    \node at (axis cs:-14, 0.05) {$p_{1}(x)$};
    \node at (axis cs:9, 0.1) {$p_{2}(x)$};
    \end{axis}
    \draw[color=gray!95, dashed] (3.38, 0) -- (3.38, 2.2);
    \draw[color=gray!95, dashed] (2.32, 0) -- (2.32, 2.2);

  \end{tikzpicture}
  %\label{fig:distributions_bias}
  %}
%
%  \centering
%  \subfloat[Devices with random errors only]{
%  \begin{tikzpicture}[scale=1]
%    \begin{axis}[
%    % Define probability distribution functions
%    declare function={
%        binom(\n,\p) = \n!/(x!*(\n-x)!)*\p^x*(1-\p)^(\n-x);
%        normal(\m,\s) = 1/(\s*sqrt(2*pi))*exp(-((x-\m)^2)/(2*\s^2));
%    },
%      every axis plot post/.append style={smooth}, % All plots: smooth, no marks
%      axis x line=bottom, % no box around the plot, only x and y axis
%      axis y line=left, % the * suppresses the arrow tips
%      enlargelimits=upper, % extend the axes a bit to the right and top
%      no markers,
%      xtick=\empty,
%      ytick=\empty,
%      samples=100,
%      clip=false,
%      domain=-28:24,
%      height=1.5in,
%      width=3in,
%      % xlabel=x,
%      ylabel={$\mathlarger p(x)$},
%      xtick={-2},
%      xticklabels={,,},
%      xtick pos=left,
%    ]
%
%    \addplot[thick, black,
%             patch, mesh,
%             patch type=quadratic spline]
%    coordinates {
%        (-19,0) (-13,0.75*0.04) (-3, 0.04) %left 3 dots
%        (8, 0.75*0.04) (19, 0.00) %right 2 dots
%    };
%
%    \addplot[thick, black] {gauss2(-2, 4.5)};
%
%    \node[right] at (axis cs:28, 0)  {\textsl{x}};
%    \node[below] at (axis cs:-1, 0) {\footnotesize \SI{60}{\celsius}};
%    \node at (axis cs:-14, 0.045) {$p_{1}(x)$};
%    \node at (axis cs:4, 0.1) {$p_{2}(x)$};
%
%    \end{axis}
%    \draw[color=gray!95, dashed] (2.74, 0) -- (2.74, 2.2);
%
%  \end{tikzpicture}
%  \label{fig:distributions_wo_bias}
%  }

  \caption{Using pdfs to model devices with systematic and random errors. Ground truth is $60\,^{\circ}\mathrm{C}$. Dashed lines are means of pdfs.}
  \label{fig:distributions}
\end{figure}

To model the behavior of devices producing noisy measurements, we associate each device $i$ with a \emph{random variable} that has a \emph{probability density function} (pdf) $p_i(x)$ such as the ones shown in Figure~\ref{fig:distributions} (the x-axis in this figure represents temperature). Random variables need not be Gaussian.\footnote{The role of Gaussians in Kalman filtering is discussed in Section~\ref{sec:discussion}.} Obtaining a measurement from device $i$ corresponds to drawing a random sample from the distribution for that device. We write \mbox{$x_i {\sim} p_i(\mu_i,\sigma_i^2)$} to denote that $x_i$ is a random variable with pdf $p_i$ whose mean and variance are $\mu_i$ and $\sigma_i^2$ respectively; following convention, we use $x_i$ to represent a random sample from this distribution as well.

Means and variances of distributions model different kinds of inaccuracies in measurements. Device $i$ is said to have a {\em systematic error} or {\em bias} in its measurements if the mean $\mu_i$ of its distribution is not equal to the actual temperature $x_c$ (in general, to the value being estimated, which is known as {\em ground truth}); otherwise, the instrument is {\em unbiased}. Figure~\ref{fig:distributions} shows pdfs for two devices that have different amounts of systematic error. The variance $\sigma_i^2$ on the other hand is a measure of the {\em random error} in the measurements. The impact of random errors can be mitigated by taking many measurements with a given device and averaging their values, but this approach will not help to reduce systematic error.

In the usual formulation of Kalman filtering, it is assumed that measuring devices do not have systematic errors. However, we do not have the luxury of taking many measurements of a given state, so we must take into account the impact of random error on a single measurement. Therefore, confidence in a device is modeled formally by the variance of the distribution associated with that device; the smaller the variance, the higher our confidence in the measurements made by the device. In Figure~\ref{fig:distributions}, the fact that we have less confidence in the first device has been illustrated by making $p_1$ more spread out than $p_2$, giving it a larger variance.

%This approach to modeling confidence may seem nonintuitive since there is no consideration of how close the measurements produced by a device are to the actual temperature of the core. In particular, a device can produce measurements that are very far off from the actual temperature of the core but as long as they fall within a narrow range of values, we would still say that we have high confidence in that device.
%
%In statistics, this is explained by making a distinction between {\em accuracy} and {\em precision}. Accuracy is a measure of how close an estimate of a quantity is to the true value of that quantity (the true value is sometimes called the \emph{ground truth}). Precision on the other hand is a measure of how close the estimates are to each other, and is defined without reference to ground truth. A metaphor that is often used to explain this difference is shooting at a bull's-eye. In this case, ground truth is provided by the center of the bull's-eye. A precise shooter is one whose shots are clustered closely together even if they may be far from the bull's-eye. In contrast, the shots of an accurate but not precise shooter would be scattered widely in a region surrounding the bull's-eye. For the problems considered in this paper, ground truth may not be available, and \emph{confidence in estimates is modeled using precision}.

The informal notion that noise should affect the two devices in ``unrelated ways'' is formalized by requiring that the corresponding random variables be \emph{uncorrelated}. This is a weaker condition than requiring them to be \emph{independent}, as explained in the Appendix~\ref{sec:basic_terms}.
Suppose we are given the measurement made by one of the devices (say $x_1$) and we have to guess what the other measurement ($x_2)$ might be. If knowing $x_1$ does not give us any new information about what $x_2$ might be, the random variables are independent. This is expressed formally by the equation
\mbox{$p(x_2|x_1) = p(x_2)$}; intuitively, knowing the value of $x_1$ does not change the pdf for the possible values of $x_2$. If the random variables are only uncorrelated, knowing $x_1$ might give us new information about $x_2$ such as restricting its possible values but the mean of $x_2|x_1$ will still be $\mu_2$. Using expectations, this can be written as \mbox{$E[x_2|x_1] = E[x_2]$}, which is equivalent to requiring that \mbox{$E[(x_1{-}\mu_1)(x_2{-}\mu_2)]$}, the covariance between the two variables, be equal to zero. This is obviously a weaker condition than independence.

Although the discussion in this section has focused on measurements, the same formalization
can be used for estimates produced by an estimator. Lemma~\ref{lemma:lc}(i)
shows how the mean and variance of a linear combination of pairwise uncorrelated
random variables can be computed from the means and variances of the random
variables~\cite{maybeck1982stochastic}. The mean and variance can be
used to quantify bias and random errors for the estimator as in the case of
measurements.

An \emph{unbiased estimator} is one whose mean is equal to the unknown value being estimated and it is preferable to a biased estimator with the same variance. Only unbiased estimators are considered in this paper. Furthermore, an unbiased estimator with a smaller variance is preferable to one with a larger variance since we would have more confidence in the estimates it produces. As a step towards generalizing this discussion to estimators that produce vector estimates, we refer to the variance of an unbiased scalar estimator as the \emph{Mean Square Error} of that estimator or {\em MSE} for short.

Lemma~\ref{lemma:lc}(ii) asserts that if a random variable is pairwise uncorrelated with a set of random variables, it is uncorrelated with any linear combination of those variables.
% https://tex.stackexchange.com/questions/219579/tikz-binomial-distribution-plus-gaussian-approximation/286198

% \begin{figure}[htb]
%     \centering
%     \includegraphics[width=0.35\textwidth,height=1.5in]{images/distributions.pdf}
%     \caption{Distributions}
%     \label{fig:distributions}
% \end{figure}

\begin{lemma}
\label{lemma:lc}
Let $x_1{\sim}p_1(\mu_1,\sigma_1^2), ...,x_n{\sim}p_n(\mu_n,\sigma_n^2)$ be a set of pairwise uncorrelated random variables. Let $y = \sum_{i=1}^n \alpha_i x_i$ be a random variable that is a linear combination of the $x_i$'s.
\begin{closeenumeratei}
\item The mean and variance of $y$ are:
\begin{align}
\label{eq1}
\mu_{y} &= \smashoperator[r]{\sum_{i=1}^n}\alpha_i \mu_i\\
\label{eq2}
\sigma_{y}^2 &= \smashoperator[r]{\sum_{i=1}^n}\alpha_i^2 \sigma_i^2
\end{align}

\item If a random variable $x_{n{+}1}$ is pairwise uncorrelated with $x_1,..,x_n$, it is uncorrelated with $y$.
\end{closeenumeratei}\item
\end{lemma}

\subsection{Vector estimates}

In some applications, estimates are vectors. For example, the state of a mobile robot might be represented by a vector containing its position and velocity.
Similarly, the vital signs of a person might be represented by a vector containing his temperature, pulse rate and blood pressure. In this paper, we denote a vector by a boldfaced lowercase letter, and a matrix by an uppercase letter.

The covariance matrix $\Sigma_{\kvecsub{x}\kvecsub{x}}$ of a random variable $\kvec{x}$ is the matrix \mbox{$E[(\kvec{x}-\pmb{\mu}_{\kvecsub{x}})(\kvec{x}-\pmb{\mu}_{\kvecsub{x}})^{\rm T}]$}, where $\pmb{\mu}_{\kvecsub{x}}$ is the mean of $\kvec{x}$. Intuitively, entry $(i{,}j)$ of this matrix is the covariance between the $i$ and $j$ components of vector $\kvec{x}$; in particular, entry $(i{,}i)$ is the variance of the $i^{th}$ component of $\kvec{x}$. A random variable $\kvec{x}$  with a pdf $p$ whose mean is $\pmb{\mu_x}$ and covariance matrix is ${\Sigma_{\kvecsub{x}\kvecsub{x}}}$ is written as $\kvec{x} {\sim} p(\pmb{\mu_x},\Sigma_{\kvecsub{x}\kvecsub{x}})$. The inverse of the covariance matrix (${\Sigma}_{\kvecsub{x}\kvecsub{x}}^{-1}$) is called the precision or \emph{information} matrix.

\emph{Uncorrelated random variables}: The cross-covariance matrix $\Sigma_{\kvecsub{v}\kvecsub{w}}$ of two random variables
$\kvec{v}$ and $\kvec{w}$ is the matrix \mbox{$E[(\kvec{v} {-} \pmb{\mu}_{\rm v})(\kvec{w} {-}\pmb{\mu}_w)^{\rm T}]$}. Intuitively, element $(i{,}j)$ of this matrix is the covariance between elements $\kvec{v}(i)$ and $\kvec{w}(j)$. If the random variables are uncorrelated,
all entries in this matrix are zero, which is equivalent to saying that every component of
$\kvec{v}$ is uncorrelated with every component of $\kvec{w}$. Lemma~\ref{lemma:mse} generalizes Lemma~\ref{lemma:lc}.

\begin{lemma}
\label{lemma:mse}
Let $\kvec{x}_1{\sim} p_1 (\pmb{\mu}_1, {\Sigma}_1),...,\kvec{x}_n{\sim} p_n  (\pmb{\mu}_n, {\Sigma}_n)$ be a set of pairwise uncorrelated random variables of length $m$. Let \mbox{$\kvec{y} = \sum_{i=1}^n \kmat{A}_i \kvec{x}_i$}.
\begin{closeenumeratei}
\item The mean and covariance matrix of $\kvec{y}$ are the following:
\begin{align}
\label{eq3}
\pmb{\mu}_{\kvecsub{y}} &= \smashoperator[r]{\sum_{i=1}^n}\kmat{A}_i \pmb{\mu}_i \\
\label{eqn:sigma-vec}
{\Sigma}_{\kvecsub{y}\kvecsub{y}} &= \smashoperator[r]{\sum_{i=1}^n}\kmat{A}_i {\Sigma}_i \kmat{A}_i^{\rm T}
%MSE(\kvec{y})
%    &= E\big\{\smashoperator[r]{\sum_{i=1}^n} (\kvec{x}_i-\pmb{\mu}_i)^{\rm T}{\kmat{A}_i}^{\rm T} \kmat{A}_i (\kvec{x}_i-\pmb{\mu}_i) \big\}
\end{align}
\item If a random variable $\kvec{x}_{n{+}1}$ is pairwise uncorrelated with $\kvec{x}_1,..,\kvec{x}_n$, it is uncorrelated with $\kvec{y}$.
\end{closeenumeratei}
\end{lemma}

The \mse of an unbiased estimator $\kvecsub{y}$ is  $E[(\kvec{y} {-} \pmb{\mu}_{\rm y} )^{\rm T}(\kvec{y} {-}\pmb{\mu}_{\rm y})]$, which is the sum of the variances of the components
of $\kvec{y}$; if $\kvec{y}$ has length 1, this reduces to variance as expected.
The \mse is also the sum of the diagonal elements of ${\Sigma}_{\kvecsub{y}\kvecsub{y}}$
(this is called the \emph{trace} of ${\Sigma}_{\kvecsub{y}\kvecsub{y}}$).

%% file: scalar.tex
Section~\ref{sec:scalarTwo} discusses the problem of fusing two scalar
estimates. Section~\ref{sec:scalarMultiple} generalizes this to the problem
of fusing $n{>}2$ scalar estimates. Section~\ref{sec:incr} shows that fusing $n{>}2$ estimates can be done iteratively by fusing two estimates at a time without any loss of quality in the final estimate.

\subsection{Fusing two scalar estimates}
\label{sec:scalarTwo}

We now consider the problem of choosing the optimal values of the parameters $\alpha$
and $\beta$ in the linear estimator \mbox{$\beta{*}x_1 + \alpha{*}x_2$} for fusing estimates $x_1$ and $x_2$ from uncorrelated random variables.

The first reasonable requirement is that if the two estimates $x_1$ and $x_2$ are equal,
fusing them should produce the same value. This implies that $\alpha{+}\beta{=}1$. Therefore the linear estimators of interest are of the form
\begin{align}
\label{eqn:unbiasedScalar}
y_{\alpha}(x_1,x_2){=}(1{-}\alpha){*}x_1 + \alpha{*}x_2
\end{align}

If $x_1$ and $x_2$ in Equation~\ref{eqn:unbiasedScalar} are considered to be unbiased estimators of some quantity of interest, then $y_{\alpha}$ is an unbiased estimator for any value of $\alpha$. How should optimality of such an estimator be defined? One reasonable definition is that the optimal value of $\alpha$ \emph{minimizes the variance of $y_{\alpha}$} since this will produce the highest-confidence fused estimates as discussed in
Section~\ref{sec:formal}. The variance (\mse) of $y_{\alpha}$ can be determined from Lemma~\ref{lemma:lc}:
\begin{align}
\label{eqn:var}
\sigma_{y}^2(\alpha) = (1-\alpha)^2{*}\sigma_1^2 + \alpha^2{*}\sigma_2^2
\end{align}

%From Lemma~\ref{lemma:lc}, we get
%\begin{align}
%\label{eqn:var}
%\mse(y_\alpha) = \sigma_y^2(\alpha) = (1-\alpha)^2{*}\sigma{_1}{^2} + \alpha^2{*}\sigma{_2}{^2}.
%\end{align}

\begin{theorem}
\label{lemma:two}
Let $x_1{\sim} p_1(\mu_1,\sigma_1^2)$ and $x_2{\sim} p_2(\mu_2,\sigma_2^2)$ be uncorrelated random variables. Consider the linear estimator \\
$y_{\alpha}(x_1,x_2) = (1{-}\alpha){*}x_1 + \alpha{*}x_2$.
The variance of the estimator is minimized for $\alpha = \frac{\sigma_1^2}{\sigma_1^2 + \sigma_2^2}$.
\end{theorem}

This result can be proved by setting the derivative of $\sigma_y^2(\alpha)$ with respect to $\alpha$ to zero and solving equation for $\alpha$.
\iftoggle{arXiv}{

\begin{proof}
  %Differentiating $\sigma_y^2(\alpha)$ with respect to $\alpha$ and setting the derivative to zero.
  \begin{align}
      \nonumber
      \frac{\mathrm{d}}{\mathrm{d}\alpha} \sigma_{y}^2(\alpha) &= -2(1-\alpha)*\sigma^2_1 + 2\alpha*\sigma^2_2 \\
      \nonumber
      &= 2\alpha*(\sigma^2_1 + \sigma^2_2) - 2*\sigma^2_1 = 0 \\
      \alpha &= \frac{\sigma_1^2}{\sigma_1^2 + \sigma_2^2}
  \end{align}

  The second order derivative of $\sigma_y^2(\alpha)$, \mbox{($\sigma_1^2 {+} \sigma_2^2$)}, is positive,
  showing that $\sigma_y^2(\alpha)$ reaches a minimum at this point.
\end{proof}
}{}

In the literature, the optimal value of $\alpha$ is called the {\em Kalman gain} $K$. Substituting $K$ into the linear fusion model, we get the optimal linear estimator
$y(x_1,x_2)$:
\begin{align}
y(x_1,x_2) &= \frac{\sigma_2^2}{\sigma_1^2 + \sigma_2^2}{*}x_1 + \frac{\sigma_1^2}{\sigma_1^2 + \sigma_2^2}{*}x_2
\end{align}

As a step towards fusion of $n{>}2$ estimates, it is useful to rewrite this as follows:
\begin{align}
\label{eqn:8a}
y(x_1,x_2) &= \frac{\frac{1}{\sigma_1^2}}{\frac{1}{\sigma_1^2} + \frac{1}{\sigma_2^2}}{*}x_1 + \frac{\frac{1}{\sigma_2^2}}{\frac{1}{\sigma_1^2} + \frac{1}{\sigma_2^2}}{*}x_2
\end{align}

%From Lemma~\ref{lemma:lc}, we get
%\begin{align}
%\label{eqn:var}
%\sigma_y^2(\alpha) = (1-\alpha)^2{*}\sigma{_1}{^2} + \alpha^2{*}\sigma{_2}{^2}
%\end{align}

Substituting the optimal value of $\alpha$ into Equation~\ref{eqn:var}, we get
\begin{align}
\label{eqn:10}
\sigma_{y}^2 &= \frac{1}{\frac{1}{\sigma_1^2} + \frac{1}{\sigma_2^2}}
\end{align}

The expressions for $y$ and $\sigma_{y}^2$ are complicated because they contain the reciprocals of variances. If we let ${\inform}_1$ and ${\inform}_2$ denote the precisions of the two distributions, the expressions for $y$ and ${\inform}_{y}$
can be written more simply as follows:
\begin{align}
\label{eqn:9}
% y(x_1,x_2) &= \frac{\frac{1}{\sigma_1^2}}{\frac{1}{\sigma{_1}{^2}} + \frac{1}{\sigma{_2}{^2}}}{*}x_1 + \frac{\frac{1}{\sigma_1^2}}{\frac{1}{\sigma{_1}{^2}} + \frac{1}{\sigma{_2}{^2}}}{*}x_2
y(x_1,x_2) &= \frac{{\inform}_1}{{\inform}_1 + {\inform}_2}{*}x_1 + \frac{{\inform}_2}{{\inform}_1 + {\inform}_2}{*}x_2 \\
\label{eqn:9b}
{\inform}_{y} &= {\inform}_1 + {\inform}_2
\end{align}

These results say that the weight we should give to an estimate is proportional
to the confidence we have in that estimate, and that we have more
confidence in the fused estimate than in the individual estimates, which is intuitively reasonable. To use these results, we need only the variances of the distributions.
In particular, the pdfs $p_i$, which are usually not available in applications, are not needed, and the proof of Theorem~\ref{lemma:two} does not require these pdf's to have the same mean.

%Moreover the distributions do not have to have the same mean,
%and the means do not need to be equal to the unknown value being estimated. For example, estimates from instruments with different systematic errors can be fused optimally using Theorem~\ref{lemma:two} since optimality is defined in terms of MMSE.

%The precision of $y$ is $i_{y} = {\inform}_1 + {\inform}_2$.

%The value $\frac{\inform_2}{\inform_1 + \inform_2}$ is sometimes called the {\em Kalman gain} $K$.

\subsection{Fusing multiple scalar estimates}
\label{sec:scalarMultiple}

The approach in Section~\ref{sec:scalarTwo} can be generalized to optimally fuse multiple
pairwise uncorrelated estimates $x_1,x_2,...,x_n$. Let $y_{n{,}\alpha}(x_1,..,x_n)$ denote the linear estimator for fusing the $n$ estimates given parameters $\alpha_1,..,\alpha_n$, which we denote by ${\alpha}$. The notation $y_{\alpha}(x_1,x_2)$ introduced in the previous section can be considered to be an abbreviation of $y_{2{,}\alpha}(x_1,x_2)$.

\begin{theorem}
\label{th:multiple}
Let $x_i {\sim} p_i(\mu_i,\sigma_i^2)$ for $(1 {\leq} i {\leq} n)$ be a set of pairwise uncorrelated random variables. Consider the linear estimator
\mbox{$y_{n{,}\alpha}(x_1,..,x_n) = \sum_{i=1}^{n} \alpha_i x_i$} where \mbox{$\sum_{i=1}^n \alpha_i = 1$}. The variance of the estimator is minimized for
\begin{align*}
\alpha_i = \frac{\frac{1}{\sigma_i^2}}{\sum_{j=1}^n \frac{1}{\sigma_j^2}}
\end{align*}
\end{theorem}

 \iftoggle{arXiv}{
 \begin{proof}
 From Lemma~\ref{lemma:lc}, $\sigma_y^2(\alpha) = \sum_{i=1}^n {\alpha_i}^2 \sigma{_i}{^2}$. To find the values of $\alpha_i$  that minimize the variance $\sigma_y^2$ under the constraint that the $\alpha_i$'s sum to $1$, we use the method of Lagrange multipliers. Define
 \begin{align*}
 f(\alpha_1,...,\alpha_n) &= \smashoperator[r]{\sum_{i=1}^n} {\alpha_i}^2 \sigma{_i}{^2} + \lambda(\smashoperator[r]{\sum_{i=1}^n}\alpha_i -1)
 \end{align*}
 where $\lambda$ is the Lagrange multiplier. Taking the partial derivatives of $f$ with respect to each $\alpha_i$ and setting these derivatives to zero, we find
 \mbox{$\alpha_1\sigma_1^2 = \alpha_2\sigma_2^2 = ... = \alpha_n\sigma_n^2 = -\lambda/2$}.
 From this, and the fact that sum of the $\alpha_i$'s is $1$, the result follows.
 \end{proof}}{}

The minimal variance is given by the following expression:
\begin{align}
\label{eqn:13}
%\sigma^2{_{y}} &=\frac{1}{\smashoperator[r]{\sum_{i=1}^n}{\frac{1}{\sigma_i^2}}} \qual
\sigma^2_{y_n} &=\frac{1}{\smashoperator[r]{\sum_{j=1}^n}{\frac{1}{\sigma_j^2}}}
\end{align}

As in Section \ref{sec:scalarTwo}, these expressions are more intuitive if the
variance is replaced with precision: the contribution of $x_i$ to the value of
$y_n({x_1,..,x_n})$ is proportional to $x_i$'s confidence.
\begin{align}
  \label{eqn:p12a}
  y_n(x_1,..,x_n) &= \smashoperator[r]{\sum_{i=1}^n}\frac{\inform_i}{{\inform}_1{+}...{+}{\inform}_n} *x_i \\
  \label{eqn:p12b}
   {\inform}_{y_n} &= \sum_{i=1}^n{{\inform}_i}
\end{align}

Equations~\ref{eqn:p12a} and \ref{eqn:p12b} generalize Equations~\ref{eqn:9} and \ref{eqn:9b}.

\subsection{Incremental fusing is optimal}\label{sec:incr}

In many applications, the estimates $x_1,x_2,...,x_n$ become available successively
over a period of time. While it is possible to store all the estimates and use Equations~\ref{eqn:p12a} and \ref{eqn:p12b} to fuse all the estimates from scratch whenever a new estimate becomes available, it is possible to save both time and storage if one can do this fusion incrementally. We show that just as a sequence of numbers can be added by keeping a running sum and adding the numbers to this running sum one at a time, a sequence of $n {>} 2$ estimates can be fused by keeping a ``running estimate'' and fusing estimates from the sequence one at a time into this running estimate without any loss in the quality of the final estimate. In short, we want to show that \mbox{$y_n(x_1,x_2,...,x_n) = y_2(y_2(y_2(x_1,x_2),x_3)... ,x_n)$}.

A little bit of algebra shows that if $n{>}2$, Equations~\ref{eqn:p12a} and \ref{eqn:p12b} for the optimal linear estimator and its precision can be expressed as shown in Equations~\ref{eqn:p12c} and \ref{eqn:p12d}.

\begin{align}
  \label{eqn:p12c}
  y_n(x_1,..,x_n) &= \frac{\nu_{y_{n{-}1}}}{\nu_{y_{n{-}1}}{+}\nu_{n}} y_{n{-}1}(x_1,...,x_{n{-}1}) + \frac{\nu_{n}}{\nu_{y_{n{-}1}}{+}\nu_{n}}x_n\\
  \label{eqn:p12d}
   {\inform}_{y_n} &= \nu_{y_{n{-}1}} + \nu_n
\end{align}

This shows that $y_n(x_1,..,x_n) = y_2(y_{n{-}1}(x_1,..,x_{n{-}1}), x_n)$. Using this argument recursively gives the required result.\footnote{We thank Mani Chandy for showing us this approach to proving the result.}

To make the connection to Kalman filtering, it is useful to derive the same result using a pictorial argument. Figure~\ref{fig:incremental-fusion} shows the process of incrementally fusing the $n$ estimates. In this picture, time progresses from left to right, the precision of each estimate is shown in parentheses next to it, and the weights on the edges are the weights from Equation~\ref{eqn:9}. The contribution of each $x_i$ to the final value \mbox{$y_2(y_2(y_2(x_1,x_2),x_3)... ,x_n)$} is given by the product of the weights on the path from $x_i$ to the final value and this product is obviously equal to the weight of $x_i$ in Equation~\ref{eqn:p12a}, showing that incremental fusion is optimal.

%Imagine that time progresses from left to right in this picture. Estimate $x_1$ is available initially, and the other estimates $x_i$ become available in succession; the precision of each estimate is shown in parentheses next to each estimate. When the estimate $x_2$ becomes available, it is fused with $x_1$ using Equation~\ref{eqn:9}. In Figure~\ref{fig:incremental-fusion}, the labels on the edges connecting $x_1$ and $x_2$ to $y_2(x_1,x_2)$ are the weights given to these estimates in Equation~\ref{eqn:9}. When estimate $x_3$ becomes available, it is fused with $y_2(x_1,x_2)$ using Equation~\ref{eqn:9}; this is justified because by assumption, random variable $x_3$ is uncorrelated to random variables $x_1$ and $x_2$, and by Lemma~\ref{lemma:lc}(ii), it is therefore uncorrelated to $y_2(x_1,x_2)$, which is a linear combination of $x_1$ and $x_2$. This process continues until all the estimates have been fused.

%The contribution of $x_i$ to the value of \mbox{$y_2(y_2(y_2(x_1,x_2),x_3)... ,x_n)$} is given by the product of the weights on the path from $x_i$ to the final value in Figure~\ref{fig:incremental-fusion}. This product has the same value as the weight of $x_i$ in Equation~\ref{eqn:p12a}, showing that incremental fusion is optimal.
%\begin{gather*}
%\frac{{\inform}_i}{{\inform}_1+...+{\inform}_i} * \frac{{\inform}_1+...+{\inform}_i}{{\inform}_1+...+{\inform}_{i+1}} *...* \frac{{\inform}_1+...+{\inform}_{n-1}}{{\inform}_1+...+{\inform}_n} \\
%= \frac{{\inform}_i}{\inform_1+...+{\inform}_n}
%\end{gather*}

\newcommand{\tztx}[1]{\text{\tiny{#1}}}
\begin{figure}
  \centering
    \begin{tikzpicture}[scale=1.0]
      \draw [->] (0.6,3) -- (2,3);
      \draw [black] (2.2,3) circle [radius=0.1];
      \node at (2.2,3) {\textbf{\large +}};
      \draw [->] (2.2,2) -- (2.2,2.8);
      \node at (0.3,3) {\tztx{($\nu_{1}$)}};
      \node at (0.3,2.8) {\tztx{$x_{1}$}};
      \node at (1.2, 3.2) {\tztx{$\frac{\nu_1}{\nu_1 + \nu_2}$}};
      \node at (2.2, 1.8) {\tztx{$x_{2}$}};
      \node at (2.2, 1.6) {\tztx{($\nu_{2}$)}};
      \node at (1.8, 2.4) {\tztx{$\frac{\nu_2}{\nu_1+\nu_2}$}};
      \node at (2.2, 3.6) {\tztx{$y(x_1, x_2)$}};
      \node at (2.2, 3.8) {\tztx{($\nu_1 + \nu_2$)}};

      \draw [->] (2.4,3) -- (4,3);
      \draw [black] (4.2,3) circle [radius=0.1];
      \node at (4.2,3) {\textbf{\large +}};
      \draw [->] (4.2,2) -- (4.2,2.8);
      \node at (4.2, 1.8) {\tztx{$x_3$}};
      \node at (4.2, 1.6) {\tztx{($\nu_{3}$)}};
      \node at (3.2, 3.2) {\tztx{$\frac{\nu_1+\nu_2}{\nu_1+\nu_2+\nu_3}$}};
      \node at (3.6, 2.4) {\tztx{$\frac{\nu_3}{\nu_1+\nu_2+\nu_3}$}};
      \node at (5.2, 3) {$\dotso$};
      \node at (4.2, 3.6) {\tztx{$y_2(y_2(x_1,x_2),x_3)$}};
      \node at (4.2, 3.8) {\tztx{($\nu_1+\nu_2+\nu_3)$}};

      \draw [black] (6.1,3) circle [radius=0.1];
      \node at (6.1,3) {\textbf{\large +}};
      \draw [->] (6.1,2) -- (6.1,2.8);
      \node at (6.1, 1.8) {\tztx{$x_{n-1}$}};
      \node at (6.1, 1.6) {\tztx{($\nu_{n-1}$)}};
      \node at (5.4, 2.4) {\tztx{$\frac{\nu_{n-1}}{\nu_1+{\dotsb}+\nu_{n-1}}$}};
      \node at (6.1, 3.6) {\tztx{$y_2(y_2({\dotsc}), x_{n-1})$}};
      \node at (6.1, 3.8) {\tztx{($\nu_1+{\dotsb}+\nu_{n-1}$)}};

      \draw [->] (6.3,3) -- (7.6,3);
      \draw [black] (7.9,3) circle [radius=0.1];
      \node at (7.9,3) {\textbf{\large +}};
      \draw [->] (7.9, 2) -- (7.9,2.8);
      \node at (7.9, 1.8) {\tztx{$x_{n}$}};
      \node at (7.9, 1.6) {\tztx{($\nu_{n}$)}};
      \node at (7.3, 2.4) {\tztx{$\frac{\nu_n}{\nu_1+{\dotsb}+\nu_n}$}};
      \node at (7.0, 3.2) {\tztx{$\frac{\nu_1+{\dotsb}+\nu_{n-1}}{\nu_1+{\dotsb}+\nu_n}$}};
      \node at (7.9, 3.6) {\tztx{$y_2(y_2({\dotsc}), x_n)$}};
      \node at (7.9, 3.8) {\tztx{($\nu_1+{\dotsb}+\nu_n$)}};

      \draw [->] (8.1,3) -- (8.6,3);
    \end{tikzpicture}
  \caption{Dataflow graph for incremental fusion.}
  \label{fig:incremental-fusion}
\end{figure}
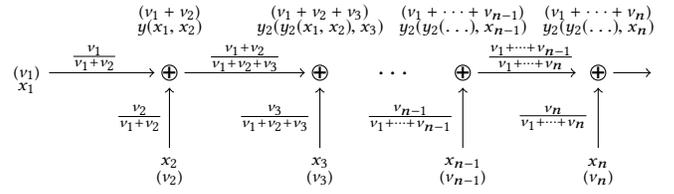

\subsection{Summary}

The results in this section can be summarized informally as follows. \emph{When using a linear estimator to fuse uncertain scalar estimates, the weight given to each estimate should be inversely proportional to the variance of the random variable from which that estimate is obtained. Furthermore, when fusing $n{>}2$ estimates, estimates can be fused incrementally without any loss in the quality of the final result.} These results are often expressed formally in terms of the Kalman gain $K$, as shown below; the equations can be applied recursively to fuse multiple estimates. Note that if $\nu_1 {\gg} \nu_2$, $K{\approx}0$ and $y(x_1{,}x_2){\approx}x_1$; conversely if $\nu_1 {\ll} \nu_2$, $K{\approx}1$ and $y(x_1{,}x_2){\approx}x_2$.

\vspace*{0.05in}

\fbox{
\begin{minipage}{3 in}
\begin{center}
    $x_1{\sim} p_1(\mu_1,\sigma_1^2),\ \  x_2{\sim}p_2(\mu_2,\sigma_2^2)$
\end{center}
\begin{align}
\label{eqn:K0}
K &= \frac{\sigma_1^2}{\sigma_1^2 + {\sigma_2^2}} = \frac{{\inform}_2}{{\inform}_1+{\inform}_2}\\
\label{eqn:K1}
y(x_1,x_2) &= x_1 + K(x_2 - x_1)\\
%\label{eqn:K2}
%\mu_{y} &= \mu_1 + K(\mu_2 - \mu_1)\\
\label{eqn:K3}
\sigma_{y}^{2} &= (1 - K)\sigma_1^2 \quad \quad \textrm{or}\quad \quad     {\inform}_{y} = {\inform}_1 + {\inform}_2
\end{align}
\end{minipage}
}

%%% Local Variables:
%%% mode: latex
%%% TeX-master: "paper"
%%% End:

%% file: vector.tex
 The results in Section~\ref{sec:scalar} for fusing scalar estimates can be extended to vectors by replacing {\em variances} with {\em covariance matrices}.

\subsection{Fusing multiple vector estimates}
\label{sec:fusemultivectorestimates}

For vectors, the linear estimator is
\begin{align}
\label{eqn:vecfusion}
\kvec{y}_{n{,}A}(\kvec{x}_1,\kvec{x}_2,..,\kvec{x}_n) = \sum_{i=1}^n \kmat{A}_i\kvec{x}_i \ \ {\rm where}\ \  \sum_{i=1}^n \kmat{A}_i = \kmat{I}
\end{align}
Here $A$ stands for the matrix parameters $(\kmat{A}_1,...,\kmat{A}_n)$. All the vectors $\kvec{x}_i$ are assumed to be of the same length. To simplify notation, we omit the subscript $n$ in $\kvec{y}_{n{,}A}$ in the discussion below since it is obvious from the context.

\emph{Optimality:} The parameters $\kmat{A}_1,...,\kmat{A}_n$ in the linear data fusion model are chosen to minimize $\mse (\kvec{y}_{A})$ which is $E[(\kvec{y}_{A} {-} \pmb{\mu}_{y_{A}} )^{\rm T}(\kvec{y}_{A} {-}\pmb{\mu}_{y_{A}})]$, as explained in Section~\ref{sec:formal}.

%The approaches introduced in Sections~\ref{sec:scalarMultiple} and \ref{sec:fusevectorestimates} can be further generalized to optimally fuse multiple vector
%estimates \kvec{x$_1$}, \kvec{x$_2$}, ..., \kvec{x$_n$}. Let \kvec{y$_A$}(\kvec{x$_1$},...,\kvec{x$_n$}) denote the linear estimator
%given parameters (\kmat{A$_1$}, ...,\kmat{A$_n$}), which we denote by \kmat{A$_i$}.
%\begin{align*}
%  \kvec{y}_\kmat{A}(\kvec{x}_1,...,\kvec{x}_n) &= \smashoperator[r]{\sum_{i=1}^n} \kmat{A}_i \kmat{x}_i \ \ \ \ {\rm where}\ \smashoperator[r]{\sum_{i=1}^n}\kmat{A}_i = I
%\end{align*}
%
%From Lemma~\ref{lemma:mse}, it follows that
%\begin{align}
%\label{eqn:mse}
%\mse(\kvec{y}) &= E\big\{\smashoperator[r]{\sum_{i=1}^n} (\kvec{x}_i-\overbar{\kvec{x}_i})^T{\kmat{A}_i}^T \kmat{A}_i (\kvec{x}_i-\overbar{\kvec{x}_i}) \big\}
%\end{align}

Theorem~\ref{th:multipleVec} generalizes Theorem~\ref{th:multiple} to the vector case.
The proof of this theorem uses matrix derivatives~\cite{matrix-cookbook} (see Appendix~\ref{sec:matrix_derivative}) and is given in Appendix~\ref{sec:proofMultipleVec} since it is not needed for understanding the rest of this paper. Comparing Theorems~\ref{th:multipleVec} and \ref{th:multiple}, we see that the expressions are similar, the main difference being that the role of variance in the scalar case is played by the covariance matrix in the vector case.

\begin{theorem}
\label{th:multipleVec}
Let $\kvec{x}_i {\sim} p_i(\pmb{\mu_i},\Sigma_i)$ for $(1 {\leq} i {\leq} n)$ be a set of pairwise uncorrelated random variables. Consider the linear estimator
\mbox{$\kvec{y}_{A}(\kvec{x}_1,..,\kvec{x}_n) {=} \sum_{i=1}^n \kmat{A}_i \kvec{x}_i$}, where \mbox{$\sum_{i=1}^n \kmat{A}_i = \kmat{I}$}. The value of $MSE(\kvec{y}_{A})$ is minimized for
\begin{align}
\label{eqn:vector_A_cov}
\kmat{A}_i &=  (\smashoperator[r]{\sum_{{\rm j}=1}^n} {\Sigma}_j^{-1})^{-1} {\Sigma}_i^{-1}
\end{align}
\end{theorem}

Therefore the optimal estimator is
\begin{align}
\label{eqn:vector_sigma}
\kvecsub{y}(\kvec{x}_1,...,\kvec{x}_n) &= ({\smashoperator[r]{\sum_{{\rm j}=1}^n}{{\Sigma}_j^{-1}}})^{-1} \smashoperator[r]{\sum_{i=1}^n} {\Sigma}_i^{-1} \kvec{x}_i
\end{align}

The covariance matrix of $\kvec{y}$ can be computed by using Lemma~\ref{lemma:mse}.
\begin{align}
\label{eqn:vector_sigma_cov}
\Sigma_{\kvecsub{y}\kvecsub{y}} &= ({\smashoperator[r]{\sum_{j=1}^n}{{\Sigma}_j^{-1}}})^{-1}
\end{align}

In the vector case, precision is the inverse of a covariance matrix, denoted by $\Inform$. Equations~\ref{eqn:vector_y_hat}--\ref{eqn:vector_P} use precision to express the optimal estimator and its variance and generalize Equations~\ref{eqn:p12a}--\ref{eqn:p12b} to the vector case.
\begin{align}
\label{eqn:vector_y_hat}
  \kvec{y}(\kvec{x}_1,...,\kvec{x}_n) &= \Inform_{\kvecsub{y}}^{{-}1}   \smashoperator[r]{\sum_{i=1}^n} \Inform_i \kvec{x}_i \\
\label{eqn:vector_P}
  \Inform_{\kvecsub{y}}&=\sum_{j=1}^n{\Inform_j}
\end{align}

% SB: We should add this proof to the arXiv report.
As in the scalar case, fusion of $n{>}2$ vector estimates can be done
incrementally without loss of precision. The proof is similar to the scalar case, and is omitted.

There are several equivalent expressions for the Kalman gain for the fusion of two estimates. The following one, which is easily derived from Equation~\ref{eqn:vector_A_cov}, is the vector analog of Equation~\ref{eqn:K0}:
\begin{align}
K &= \Sigma_1(\Sigma_1 + \Sigma_2)^{-1}
\end{align}

The covariance matrix of the optimal estimator $\kvec{y}(\kvec{x}_1,\kvec{x}_2)$
is the following.
\begin{align}
\label{eqn:vector_sigma_cov_mat}
{\Sigma}_{\kvecsub{y}\kvecsub{y}} &= \Sigma_1(\Sigma_1 + \Sigma_2)^{-1} \Sigma_2 \\
&= K \Sigma_2 = {\Sigma}_{1} - \kmat{K}{\Sigma}_{1}
\end{align}

\subsection{Summary}

The results in this section can be summarized in terms of the Kalman gain $\kmat{K}$ as follows.

\vspace*{0.05in}

\fbox{
\begin{minipage}{3 in}
\begin{center}
    $\kvec{x}_1{\sim}p_1(\pmb{\mu}_1,{\Sigma}_1),\ \  \kvec{x}_2{\sim}p_2 (\pmb{\mu}_2, {\Sigma}_2)$
\end{center}
\begin{align}
\label{eqn:M0}
%\kmat{K} &= (\Sigma_1^{-1} + \Sigma_2^{-1})^{-1} \Sigma_2^{-1} = \Sigma_1(\Sigma_1+\Sigma_2)^{-1}\\
\kmat{K} &= \Sigma_1(\Sigma_1+\Sigma_2)^{-1} =(\Inform_1 + \Inform_2)^{-1} \Inform_2 \\
\label{eqn:M1}
\kvec{y}(\kvec{x}_1,\kvec{x}_2) &= \kvec{x}_1 + \kmat{K}(\kvec{x}_2 - \kvec{x}_1)\\
%\label{eqn:M2}
%\pmb{\mu}_{y} &= \pmb{\mu}_1 + \kmat{K}(\pmb{\mu}_2 - \pmb{\mu}_1)\\
\label{eqn:M3}
{\Sigma}_{\kvecsub{y}\kvecsub{y}} &= (I - \kmat{K}){\Sigma}_{1}  \quad \textrm{or} \quad \Inform_{\kvecsub{y}} = \Inform_1 + \Inform_2
\end{align}
\end{minipage}
}

\iffalse
\begin{align}
%\nonumber % Remove numbering (before each equation)
  K &= \frac{\Sigma_0}{\Sigma_0 + \Sigma_1}\\
  \mu_1' &= \mu_0 + K(\mu_1 - \mu_0) \\
  \Sigma\nolimits ' &= \Sigma\nolimits_0 - K\Sigma\nolimits_0
\end{align}

\begin{align}
%\nonumber % Remove numbering (before each equation)
 A_i = (\smashoperator[r]{\sum_{{\rm j}=1}^n} \Pi_{\rm j})^{-1}\Pi_i
\end{align}
\fi

%%% Local Variables:
%%% mode: latex
%%% TeX-master: "paper"
%%% End:

%% file: hidden.tex
% say that in some problems, you have partial observability
% if so, you have to estimate the hidden state from an observation of the visible state
% this sounds impossible but if the hidden state is correlated to the visible state,
% you can use the correlation to estimate the hidden state
% we will use the Best Linear Unbiased Estimator (BLUE) for this
% define BLUE (random variable y will be estimated as Ax + b, where A is another random variable)
% and use the proof we have to find the optimal values of A and b
% find the covariance matrix of y
% mention that BLUE is not MVUE always but if y and x are jointly normally distributed,
% BLUE is MVUE

In some applications, the state of the system is represented by a vector
but only part of the state can be measured directly, so it is necessary to estimate the hidden portion of the state corresponding to a measured value of the visible
state. This section describes an estimator called the \emph{Best Linear Unbiased
Estimator} (BLUE)~\cite{mendel1995lessons, sengijpta1995fundamentals,
kitanidis1987unbiased} for doing this.

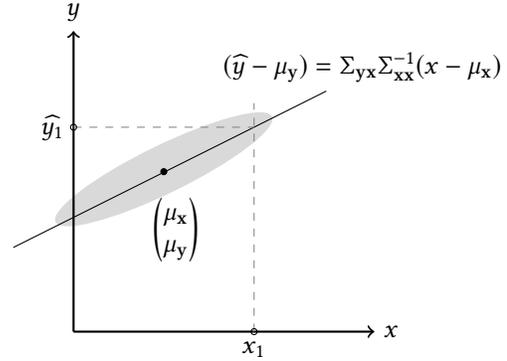
\begin{figure}[htbp]
  \centering
\begin{tikzpicture}[scale=0.8]
% \draw[help lines, color=gray!30, dashed] (0,0) grid (4.9,4.9);
 \fill [color=gray!30] (1.5,2.7) circle [x radius=20mm, y radius=4mm, rotate=26];
\draw[->, thick] (0,0)--(5,0) node[right]{$x$};
\draw[->, thick] (0,0)--(0,5) node[above]{$y$};

\draw (-1, 1.4) -- (4.2,4);
\draw[color=gray!95, dashed] (3,0) -- (3,3.8);
\draw[color=gray!95, dashed] (0,3.4) -- (3,3.4);
\draw [] (3,0) circle [radius=0.05] node[below]{$x_1$};
\draw [] (0,3.4) circle [radius=0.05] node[left]{$\widehat{y_1}$};
\draw [fill=black] (1.5,2.66) circle [radius=0.05];
\node at (1.7,1.7) {$\begin{pmatrix} \mu_{\rm {\bf x}} \\ \mu_{\rm {\bf y}} \end{pmatrix}$};
\node at (4.8,4.4) {$(\widehat{y} - \mu_{\rm {\bf y}}) = \Sigma_{\rm {\bf yx}}\Sigma_{\rm {\bf xx}}^{-1}({x}-\mu_{\rm {\bf x}})$};
\end{tikzpicture}
\caption{BLUE line corresponding to Equation~\ref{eqn:blue1}.}
\label{fig:blue}
\end{figure}

Consider the general problem of determining a value for vector $\kvec{y}$ given a
value for a vector $\kvec{x}$. If there is a functional relationship between $\kvec{x}$ and $\kvec{y}$ (say $\kvec{y} {=} F(\kvec{x})$ and $F$ is given), it is easy to compute $\kvec{y}$ given a value for $\kvec{x}$ (say $\kvec{x}_1$).

In our context however, $\kvec{x}$ and $\kvec{y}$ are random variables, so such a precise functional relationship will not hold. Figure~\ref{fig:blue} shows an example in which $x$ and $y$ are scalar-valued random variables. The gray ellipse in this figure, called a {\em confidence ellipse}, is a projection of the joint distribution of $x$ and $y$ onto the $(x,y)$ plane, that shows where some large proportion of the $(x,y)$ values are likely to be. Suppose $x$ takes the value $x_1$.
Even within the confidence ellipse, there are many points $(x_1,y)$, so we cannot associate a single value of $y$ with $x_1$. One possibility is to compute the mean of the $y$ values associated with $x_1$ (that is, the expectation \mbox{$E[y|x{=}x_1]$}), and return this as the estimate for $y$ if $x{=}x_1$. This requires knowing the joint distribution of $x$ and $y$, which may not always be available.

In some problems, we can assume that there is an unknown linear relationship between $\kvec{x}$ and $\kvec{y}$ and that uncertainty comes from noise. Therefore, we can use a technique similar to the ordinary least squares (OLS) method to estimate this linear relationship, and use it to return the best estimate of $y$ for any given value of $x$.
In Figure~\ref{fig:blue}, we see that although there are many points $(x_1,y)$, the $y$ values are clustered around the line shown in the figure so the value $\widehat{y_1}$ is a reasonable estimate for the value of $y$ corresponding to $x_1$. This line, called the \emph{best linear unbiased estimator} (BLUE), is the analog of ordinary least squares (OLS) for distributions.

\paragraph*{Computing BLUE}
\label{sec:blue}

%Let $\kvec{x}{:}p_{\rm {\bf x}}{\sim}(\pmb{\mu}_{\rm {\bf x}},\Sigma_{\rm {\bf x}{\bf x}})$ and $\kvec{y}{:}p_{\rm {\bf y}}{\sim}(\pmb{\mu}_{\rm {\bf y}},\Sigma_{\rm {\bf y}{\bf y}})$ be random variables.

Consider the estimator $\widehat{\kvec{y}}_{A,\bf{b}}(\kvec{x}) {=} A\kvec{x}{+}\kvec{b}$. We choose $A$ and $\kvec{b}$ so that this is an unbiased estimator with minimal \mse. The ``$\hspace{5pt}\widehat{}\hspace{5pt}$'' over the $\kvec{y}$ is notation that indicates that we are computing an estimate for $\kvec{y}$.

\begin{theorem}
\label{lemma:blue}
Let $\begin{pmatrix} \kvec{x} \\ \kvec{y} \end{pmatrix} {\sim}
p(\begin{pmatrix} \pmb{\mu}_\kvecsub{x} \\ \pmb{\mu}_\kvecsub{y} \end{pmatrix},
\begin{pmatrix} \Sigma_{\kvecsub{xx}}\hspace{5pt}\Sigma_{\kvecsub{xy}} \\ \Sigma_{\kvecsub{yx}}\hspace{5pt}\Sigma_{\kvecsub{yy}} \end{pmatrix})$.
The estimator $\widehat{\kvec{y}}_{A,\kvecsub{b}}(\kvec{x}) {=} A\kvec{x}{+}\kvec{b}$
for estimating the value of $\kvec{y}$ for a given value of $\kvec{x}$ is an unbiased
estimator with minimal \mse if
\begin{align*}
%\label{eqn:best_b}
\kvec{b} &= \pmb{\mu}_{\rm {\bf y}} {-} A(\pmb{\mu}_{\rm {\bf x}}) \\
%\label{eqn:best_A}
\kmat{A} &= \Sigma^{ }_{\kvecsub{yx}} \Sigma^{-1}_{\kvecsub{xx}}
\end{align*}
\end{theorem}

The proof of Theorem~\ref{lemma:blue} is straightforward. For an unbiased estimator, \mbox{$E[\widehat{\kvec{y}}] {=} E[\kvec{y}]$}. This implies that \mbox{$\kvec{b} {=} \pmb{\mu}_{\kvecsub{y}} {-} A(\pmb{\mu}_{\kvecsub{x}})$} so an unbiased estimator is of the form \mbox{$\widehat{\kvec{y}}_{A}(\kvec{x}) = \pmb{\mu}_{\kvecsub{y}} {+} A(\kvec{x} {-} \pmb{\mu}_{\kvecsub{x}})$}. Note that this is equivalent to asserting that the BLUE line must pass through the point $(\pmb{\mu}_{\kvecsub{x}}, \pmb{\mu}_{\kvecsub{y}})$. Setting the derivative of $MSE_{A}(\widehat{\kvec{y}}_A)$
with respect to $A$ to zero\cite{matrix-cookbook} and solving for $A$, we find that the best linear unbiased estimator is
\begin{align}
\label{eqn:blue1}
\widehat{\kvec{y}} &= \pmb{\mu}_{\kvecsub{y}} + \Sigma_{\kvecsub{yx}}\Sigma_{\kvecsub{xx}}^{-1}(\kvec{x} - \pmb{\mu}_{\kvecsub{x}})
\end{align}

%\begin{align*}
%MSE_{A,{\rm {\bf b}}}(\widehat{\kvec{y}}) &= E[(\kvec{y}-\widehat{\kvec{y}})^{\rm T} (\kvec{y}-\widehat{\kvec{y}})] \\
%&= E[(\kvec{y}-(A\kvec{x}+\kvec{b}))^{\rm T} (\kvec{y}-(A\kvec{x}+\kvec{b}))] \\
%&= E[\kvec{y}^{\rm T}\kvec{y}  - 2 \kvec{y}^{\rm T}(A\kvec{x}+\kvec{b}) +(A\kvec{x}+\kvec{b})^{\rm T}(A\kvec{x}+\kvec{b})]
%\end{align*}

This equation can be understood intuitively as follows. If we have no information about
$\kvec{x}$ and $\kvec{y}$, the best we can do is the estimate
\mbox{$(\pmb{\mu}_{\kvecsub{x}}, \pmb{\mu}_{\kvecsub{y}})$}, which lies on the BLUE line.
However, if we know that $\kvec{x}$ has a particular value $\kvec{x}_1$, we can use the correlation between $\kvec{y}$ and $\kvec{x}$ to estimate a better value for $\kvec{y}$
from the difference \mbox{$(\kvec{x}_1 {-} \pmb{\mu}_{\kvecsub{x}})$}.
Note that if $\Sigma_{\rm {\bf yx}} = 0$ (that is,
$\kvec{x}$ and $\kvec{y}$ are uncorrelated), the best estimate of $\kvec{y}$ is
just $\pmb{\mu}_{\kvecsub{y}}$, so knowing the value of $\kvec{x}$ does not give
us any additional information about $\kvec{y}$ as one would expect.
In Figure~\ref{fig:blue}, this corresponds to the case when the BLUE line is parallel to the x-axis. At the other extreme, suppose that $\kvec{y}$ and $\kvec{x}$ are functionally related so $\kvec{y} =  C\kvec{x}$. In that case, it is easy to see that \mbox{$\Sigma_{\kvecsub{yx}} = C \Sigma_{\kvecsub{xx}}$}, so $\est{\kvec{y}} = C \kvec{x}$ as expected. In Figure~\ref{fig:blue}, this corresponds to the case when the confidence ellipse shrinks down to the BLUE line.

Equation~\ref{eqn:blue1} is a generalization of ordinary least squares in the sense that if we compute the relevant means and variances of a set of discrete data $(x_i,y_i)$ and substitute into Equation~\ref{eqn:blue1}, we get the same line that is obtained by using OLS.

%%% Local Variables:
%%% mode: latex
%%% TeX-master: "paper"
%%% End:

%% file: dynamics.tex
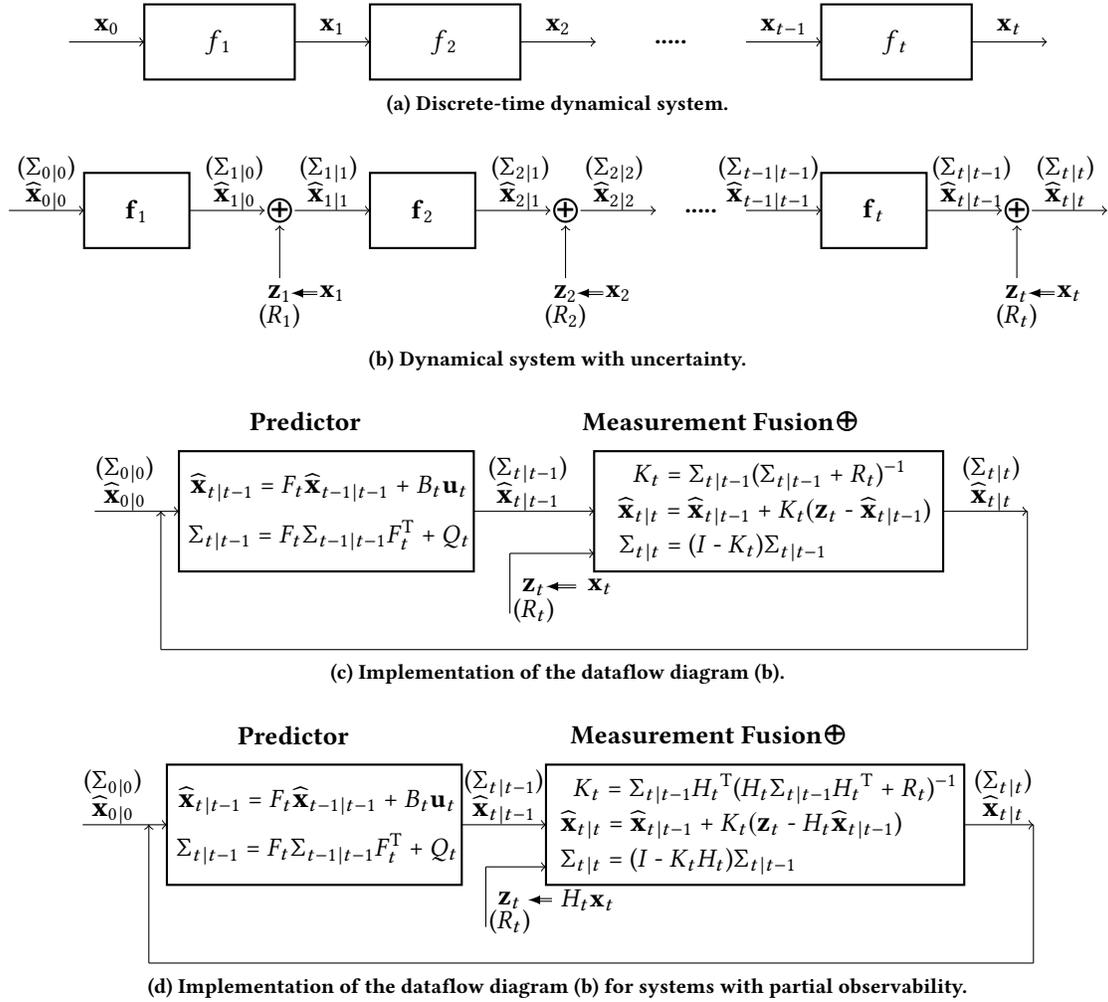
\begin{figure*}[tbhp]
  \centering
  \subfloat[Discrete-time dynamical system.]{
    \begin{tikzpicture}[scale=1.0]
      \draw [->] (0.0, 5.5) -- (1.0, 5.5);
      \draw [thick] (1.0, 5) rectangle (3.0, 6);
      \draw [->] (3.0, 5.5) -- (4.0, 5.5);
      \draw [thick] (4.0, 5) rectangle (6.0, 6);
      \draw [->] (6.0, 5.5) -- (7.0, 5.5);

      \node at (0.5, 5.7) {\textbf{x}$_{0}$};
      \node at (3.5, 5.7) {\textbf{x}$_{1}$};
      \node at (6.5, 5.7) {\textbf{x}$_{2}$};

      \node at (2, 5.5) {\textit{f}$_{1}$};
      \node at (5, 5.5) {\textit{f}$_{2}$};

      \node at (8,5.5) {\textbf{.....}};

      \draw [->] (9, 5.5) -- (10, 5.5);
      \draw [thick] (10, 5) rectangle (12, 6);
      \draw [->] (12, 5.5) -- (13, 5.5);

      \node at (9.5, 5.7) {\textbf{x}$_{t-1}$};
      \node at (12.5, 5.7) {\textbf{x}$_{t}$};
      \node at (11, 5.5) {\textit{f}$_{t}$};
    \end{tikzpicture}
    \label{fig:discrete-dynamic}
  } \\
  \subfloat[Dynamical system with uncertainty.]{
    \begin{tikzpicture}[scale=1.0]
      \draw [->] (0.0, 5.5) -- (1, 5.5);
      \draw [thick] (1, 5) rectangle (2.4, 6);
      \draw [->] (2.4, 5.5) -- (3.4, 5.5);
      \draw [black, thick] (3.6, 5.5) circle [radius=0.15];
      \draw [->] (3.6, 4.6) -- (3.6, 5.3);
      \node at (3.6, 5.5) {\textbf{\LARGE +}};
      \draw [->] (3.8, 5.5) -- (4.8, 5.5);

      \draw [thick] (4.8,5) rectangle (6.2,6);
      \draw [->] (6.2, 5.5) -- (7.2, 5.5);
      \draw [black, thick] (7.4, 5.5) circle [radius=0.15];
      \draw [->] (7.4, 4.6) -- (7.4, 5.3);
      \node at (7.4, 5.5) {\textbf{\LARGE +}};
      \draw [->] (7.6, 5.5) -- (8.6, 5.5);

      \node at (0.5, 5.7) {$\est{\kvec{x}}_{0|0}$};
      \node at (0.5, 6.05) {($\Sigma_{0|0}$)};
      \node at (1.7, 5.5) {\textbf{f}$_{1}$};
      \node at (3.0, 5.7) {$\est{\kvec{x}}_{1|0}$};
      \node at (3.0, 6.05) {($\Sigma_{1|0}$)};
      \node at (4.25, 5.7) {$\est{\kvec{x}}_{1|1}$};
      \node at (4.25, 6.05) {($\Sigma_{1|1}$)};
      \node at (3.6, 4.4) {$\kvec{z}_{1}$};
      \node at (3.6, 4.1) {($R_{1}$)};
      \draw [>=latex, ->, double] (4.1, 4.4) -- (3.75, 4.4);
      \node at (4.3, 4.4) {\textbf{x}$_{1}$};
      \node at (5.5, 5.5) {\textbf{f}$_{2}$};
      \node at (6.8, 5.7) {$\est{\kvec{x}}_{2|1}$};
      \node at (6.8, 6.05) {($\Sigma_{2|1}$)};
      \node at (7.4, 4.4) {$\kvec{z}_{2}$};
      \node at (7.4, 4.1) {($R_{2}$)};
      \draw [>=latex, ->, double] (7.9, 4.4) -- (7.55, 4.4);
      \node at (8.1, 4.4) {\textbf{x}$_{2}$};
      \node at (8.05, 5.7) {$\est{\kvec{x}}_{2|2}$};
      \node at (8.05, 6.05) {($\Sigma_{2|2}$)};

      \node at (9.2, 5.5) {\textbf{.....}};

      \draw [->] (9.8, 5.5) -- (10.8, 5.5);
      \draw [thick] (10.8, 5) rectangle (12.2, 6);
      \draw [->] (12.2, 5.5) -- (13.2, 5.5);
      \draw [black, thick] (13.4, 5.5) circle [radius=0.15];
      \draw [->] (13.6, 5.5) -- (14.6, 5.5);
      \draw [->] (13.4, 4.6) -- (13.4, 5.3);
      \node at (13.4, 5.5) {\textbf{\LARGE +}};

      \node at (10.1, 5.7) {$\est{\kvec{x}}_{t-1|t-1}$};
      \node at (10.1, 6.05) {($\Sigma_{t-1|t-1}$)};
      \node at (11.5, 5.5) {\textbf{f}$_{t}$};
      \node at (12.8, 5.7) {$\est{\kvec{x}}_{t|t-1}$};
      \node at (12.8, 6.05) {($\Sigma_{t|t-1}$)};
      \node at (14.05, 5.7) {$\est{\kvec{x}}_{t|t}$};
      \node at (14.05, 6.05) {($\Sigma_{t|t}$)};
      \node at (13.4, 4.4) {$\kvec{z}_{t}$};
      \node at (13.4, 4.1) {($R_{t}$)};
      \draw [>=latex, ->, double] (13.9, 4.4) -- (13.55, 4.4);
      \node at (14.1, 4.4) {\textbf{x}$_{t}$};
    \end{tikzpicture}
    \label{fig:dynamic-uncertainty}
  } \\

  % \subfloat[Here is the caption for the first subfigure. It should be possible to redo these figures in TikZ.]{\includegraphics[width=0.5\textwidth,height=2in]{images/dynamics.pdf}} \\

  \subfloat[Implementation of the dataflow diagram (b).]{

      \begin{tikzpicture}[scale=0.8]
      \draw [->] (0.5, 6) -- (1.9, 6);
      \draw [thick] (1.9, 5) rectangle (6.8, 7);
      \draw [->] (6.8, 6) -- (8.8, 6);
      \draw [thick] (8.8, 5) rectangle (14.6, 7);
      \draw [->] (14.6, 6) -- (16, 6);

      \node at (1, 6.7) {($\Sigma_{0|0}$)};
      \node at (1, 6.25) {$\est{\kvec{x}}_{0|0}$};

      \node at (4.4, 6.4) {$\est{\kvec{x}}_{t|t-1}$ = $F_{t}\est{\kvec{x}}_{t-1|t-1}$ + $B_{t}\textbf{u}_{t}$};
      \node at (4.4, 5.6) {$\Sigma_{t|t-1}$ = $F_{t}$$\Sigma_{t-1|t-1}$$F_{t}^{\rm T}$ + $Q_{t}$};

      \node at (7.7, 6.7)  {($\Sigma_{t|t-1}$)};
      \node at (7.7, 6.25) {$\est{\kvec{x}}_{t|t-1}$};
      \node at (11.72, 6.6) {$K_{t}$ = $\Sigma_{t|t-1}$($\Sigma_{t|t-1}$ + $R_{t}$)$^{-1}$};
      \node at (11.8, 6.0) {$\est{\kvec{x}}_{t|t}$ = $\est{\kvec{x}}_{t|t-1}$ + $K_{t}(\kvec{z}_{t}$ - $\est{\kvec{x}}_{t|t-1}$)};
      \node at (10.94, 5.4) {$\Sigma_{t|t}$ = ($I$ - $K_{t}$)$\Sigma_{t|t-1}$};

      \draw [->] (7.4, 5.3) -- (8.8, 5.3);
      \draw (7.4, 4.3) -- (7.4, 5.3);
      \node at (7.8, 4.75) {$\kvec{z}_{t}$};
      \node at (7.8, 4.35) {($R_{t}$)};
      \draw [>=latex, ->, double] (8.5, 4.75) -- (8.0, 4.75);
      \node at (8.9, 4.75) {\textbf{x}$_{t}$};

      \draw (16, 6) -- (16, 3.7);
      \draw (16, 3.7) -- (1.6, 3.7);
      \draw [->] (1.6, 3.7) -- (1.6, 6);

      \node at (15.4, 6.7) {($\Sigma_{t|t}$)};
      \node at (15.4, 6.25) {$\est{\kvec{x}}_{t|t}$};

      \node at (4, 7.5) {\textbf{Predictor}};
      \node at (10.7, 7.5) {\textbf{Measurement Fusion}};
      \draw [black, thick] (13, 7.5) circle [radius=0.15];
      \node at (13, 7.5) {\textbf{\LARGE +}};
    \end{tikzpicture}
    \label{fig:dataflow-diag}
    }

% \subfloat[Implementation of the dataflow diagram (b) for systems with partial observability.]{
\subfloat[Implementation of the dataflow diagram (b) for systems with partial observability.]{

      \begin{tikzpicture}[scale=0.8]
      \draw [->] (0.5, 6) -- (1.9, 6);
      \draw [thick] (1.9, 5) rectangle (6.8, 7);
      \draw [->] (6.8, 6) -- (8.2, 6);
      \draw [thick] (8.2, 5) rectangle (15.15, 7);
      \draw [->] (15.15, 6) -- (16.3, 6);

      \node at (1, 6.7) {($\Sigma_{0|0}$)};
      \node at (1, 6.25) {$\est{\kvec{x}}_{0|0}$};

      \node at (4.4, 6.4) {$\est{\kvec{x}}_{t|t-1}$ = $F_{t}\est{\kvec{x}}_{t-1|t-1}$ + $B_{t}$\textbf{u}$_{t}$};
      \node at (4.4, 5.6) {$\Sigma_{t|t-1}$ = $F_{t}$$\Sigma_{t-1|t-1}$$F_{t}^{\rm T}$ + $Q_{t}$};

      \node at (7.5, 6.7)  {($\Sigma_{t|t-1}$)};
      \node at (7.5, 6.25) {$\est{\kvec{x}}_{t|t-1}$};
      \node at (11.85, 6.6) {$K_{t}$ = $\Sigma_{t|t-1}H_t$$^{\rm T}$($H_t\Sigma_{t|t-1}H_t$$^{\rm T}$ + $R_{t}$)$^{-1}$};
      \node at (11.28, 6.0) {$\est{\kvec{x}}_{t|t}$ = $\est{\kvec{x}}_{t|t-1}$ + $K_{t}(\kvec{z}_{t}$ - $H_t\est{\kvec{x}}_{t|t-1}$)};
      \node at (10.4, 5.4) {$\Sigma_{t|t}$ = ($I$ - $K_{t}H_t$)$\Sigma_{t|t-1}$};

      \draw [->] (7.2, 5.3) -- (8.2, 5.3);
      \draw (7.2, 4.3) -- (7.2, 5.3);
      \node at (7.6, 4.75) {$\kvec{z}_{t}$};
      \node at (7.6, 4.4) {($R_{t}$)};
      \draw [>=latex, ->, double] (8.3, 4.75) -- (7.9, 4.75);
      \node at (8.9, 4.75) {$H_t$\textbf{x}$_{t}$};

      \draw (16.3, 6) -- (16.3, 3.7);
      \draw (16.3, 3.7) -- (1.6, 3.7);
      \draw [->] (1.6, 3.7) -- (1.6, 6);

      \node at (15.8, 6.7) {($\Sigma_{t|t}$)};
      \node at (15.8, 6.25) {$\est{\kvec{x}}_{t|t}$};

      \node at (4, 7.5) {\textbf{Predictor}};
      \node at (10.7, 7.5) {\textbf{Measurement Fusion}};
      \draw [black, thick] (13, 7.5) circle [radius=0.15];
      \node at (13, 7.5) {\textbf{\LARGE +}};
    \end{tikzpicture}
    \label{fig:dataflow-impl}
  }

    \caption{State estimation using Kalman filtering.}
    \label{fig:dynamics}
\end{figure*}

In this section, we apply the algorithms developed in Sections~\ref{sec:scalar}-\ref{sec:hidden} to the particular problem of state estimation in linear systems, which is the classical application of Kalman filtering.

Figure~\ref{fig:discrete-dynamic} shows how the evolution of the state
of such a system over time can be computed if the initial state $\kvec{x}_0$ and the model
of the system dynamics are known precisely. Time advances in discrete steps. The
state of the system at any time step is a function of the state of the system at
the previous time step and the control inputs applied to the system during that
interval. This is usually expressed by an equation of the form
\mbox{$\kvec{x}_{t} = f_t(\kvec{x}_{t-1},\kvec{u}_t)$} where $\kvec{u}_t$ is the
control input. The function $f_t$ is nonlinear in the general case, and can be
different for different steps. If the system is linear, the relation for state
evolution over time can be written as
\mbox{$\kvec{x}_{t} = \kmat{F}_t\kvec{x}_{t-1} + \kmat{B}_t\kvec{u}_t$}, where
$\kmat{F}_t$ and $\kmat{B}_t$ are time-dependent matrices that can be
determined from the physics of the system. Therefore, if the initial state
$\kvec{x}_0$ is known exactly and the system dynamics are modeled perfectly by the $F_t$ and $B_t$ matrices, the
evolution of the state over time can be computed precisely as shown in Figure~\ref{fig:discrete-dynamic}.

In general however, we may not know the initial state exactly, and the system dynamics and control inputs may not be known precisely. These inaccuracies may cause the state computed by the model to diverge unacceptably from the actual state over time. To avoid this, we can make measurements of the state after each time step. If these measurements were exact, there would of course be no need to model the system dynamics. However, in general, the measurements themselves are imprecise.

Kalman filtering was invented to solve the problem of state estimation in such systems. Figure~\ref{fig:dynamic-uncertainty} shows the dataflow of the computation, and we use it to introduce standard terminology. An estimate of the initial state, denoted by $\est{\kvec{x}}_{0|0}$, is assumed to be available. At each time step $t {=} 1,2,..$, the system model is used to provide an estimate of the state at time $t$ using information from time $t{-}1$. This step is called {\em prediction} and the estimate that it provides is called the {\em a priori} estimate and denoted by $\est{\kvec{x}}_{t|t-1}$.  The {\em a priori} estimate is then fused with $\kvec{z}_t$, the state estimate obtained from the measurement at time $t$, and the result is the {\em a posteriori} state estimate at time $t$, denoted by $\est{\kvec{x}}_{t|t}$. This {\em a posteriori} estimate is used by the model to produce the {\em a priori} estimate for the next time step and so on. As described below, the {\em a priori} and {\em a posteriori} estimates are the means of certain random variables; the covariance matrices of these random variables are shown within parentheses above each estimate in Figure~\ref{fig:dynamic-uncertainty}, and these are used to weight estimates when fusing them.

Section~\ref{sec:state} presents the state evolution model and {\em a priori} state estimation. Section~\ref{sec:Kalman-full} discusses how state estimates are fused if
an estimate of the entire state can be obtained by measurement; Section~\ref{sec:Kalman-partial} addresses this problem when only a portion of the state can be measured directly.

\subsection{State evolution model and prediction}
\label{sec:state}

The evolution of the state over time is described by a series of random variables $\kvec{x}_{0}$, $\kvec{x}_{1}$, $\kvec{x}_{2}$,...

\begin{itemize}
\item The random variable \mbox{$\kvec{x}_{0}$} captures the likelihood of different initial states.
%\item Intuitively, our belief about the initial state \mbox{$\kvec{x}_{0}$}
%    is represented by a random variable $\kvec{x}_{0|0}$, with estimated
%    mean $\est{\kvec{x}}_{0|0}$ and covariance matrix $\Sigma_{0|0}$.

\item The random variables at successive time steps are related by the following linear model:
\begin{align}
\label{eqn:state_evolution}
\kvec{x}_{t} &= \kmat{F}_t\kvec{x}_{t-1} + \kmat{B}_t\kvec{u}_t + \kvec{w}_t
\end{align}

Here, $\kvec{u}_t$ is the control input, which is assumed to be deterministic, and $\kvec{w}_t$ is a zero-mean noise term that models all the uncertainty in the system. The covariance matrix of $\kvec{w}_t$ is denoted by $\kmat{Q}_t$, and the noise terms in different time steps are assumed to be uncorrelated to each other (that is, $E[\kvec{w}_i \kvec{w}_j]{=}0$ if $i{\neq}j$) and to $\kvec{x}_{0}$.
\end{itemize}

%For estimation, we have two random variables $\kvec{x}_{t|t{-}1}$ and $\kvec{x}_{t|t}$ at each time step $t=1,2,...$. These capture beliefs about the likelihood of different states at time $t$ before and after fusion with the measurement respectively. The mean and covariance matrix of a random variable $\kvec{x}_{i|j}$ are denoted by $\est{\kvec{x}}_{i|j}$ and $\Sigma_{i|j}$ respectively.
For estimation, we have a random variable $\kvec{x}_{0|0}$ that captures {\bf our belief} about the likelihood of different states at time $t{=}0$, and two random variables $\kvec{x}_{t|t{-}1}$ and $\kvec{x}_{t|t}$ at each time step $t=1,2,...$ that capture our beliefs about
the likelihood of different states at time $t$ before and after fusion with
the measurement respectively. The mean and covariance matrix of a random variable
$\kvec{x}_{i|j}$ are denoted by $\est{\kvec{x}}_{i|j}$ and $\Sigma_{i|j}$ respectively.
We assume $E[\est{\kvec{x}}_{0|0}] {=}E[\kvec{x}_0]$ (no bias).

Prediction essentially uses $\kvec{x}_{t{-}1|t{-}1}$ as a proxy for $\kvec{x}_{t-1}$ in Equation~\ref{eqn:state_evolution} to determine $\kvec{x}_{t|t{-}1}$ as shown in Equation~\ref{eqn:state_evolution_priori}.
%(by convention, $\kvec{x}_{0|0}$ is $\kvec{x}_0$).

\begin{align}
\label{eqn:state_evolution_priori}
\kvec{x}_{t|t{-}1} &= \kmat{F}_t\kvec{x}_{t{-}1|t{-}1} + \kmat{B}_t\kvec{u}_t + \kvec{w}_t
\end{align}

For state estimation, we need only the mean and covariance matrix of $\kvec{x}_{t|t{-}1}$. The Predictor box in Figure~\ref{fig:dynamics} computes these values; the covariance matrix is obtained from Lemma~\ref{lemma:mse} under the assumption that $\kvec{w}_t$ is uncorrelated with $\kvec{x}_{t{-}1|t{-}1}$, which is justified in Section~\ref{sec:Kalman-full}.

\subsection{Fusing complete observations of the state}
\label{sec:Kalman-full}
%Full observability means that $\kvec{z}_{t+1}$ contains information about
%every state. To be more precise, the rank of $\kmat{H}_{t+1}$ should equal
%to the length of the state vector $\kvec{x}_{t+1|t}$. Full observability infers
%that $\kvec{z}_{t+1}$ and $\kvec{x}_{t+1|t}$ are of the same length.

If the entire state can be measured at each time step, the imprecise measurement at time $t$ is modeled as follows:
\begin{align}
\label{eqn:observation_eqn}
  \kvec{z}_{t} = \kvec{x}_{t} + \kvec{v}_{t}
\end{align}
\noindent
where $\kvec{v}_{t}$ is a zero-mean noise term with covariance matrix $R_{t}$. The noise terms in different time steps are assumed to be uncorrelated with each other (that is, $E[\kvec{v}_i \kvec{v}_j]$ is zero if $i{\neq}j$) as well as with $\kvec{x}_{0|0}$ and all $\kvec{w}_k$. A subtle point here is that $\kvec{x}_{t}$ in this equation is the actual state of the system at time $t$ (that is, a particular realization of the random variable $\kvec{x}_{t}$), so variability in $\kvec{z}_{t}$ comes only from $\kvec{v}_{t}$ and its covariance matrix $R_t$.

Therefore, we have two imprecise estimates for the state at each time step $t=1,2,...$,
the {\em a priori} estimate from the predictor ($\est{\kvec{x}}_{t|t{-}1}$) and the one from the measurement ($\kvec{z}_t$). If $\kvec{z}_t$ is uncorrelated to
$\kvec{x}_{t|t{-}1}$, we can use Equations~\ref{eqn:M0}-\ref{eqn:M3} to fuse the estimates as shown in Figure~\ref{fig:dataflow-diag}.

The assumptions that (i) $\kvec{x}_{t{-}1|t{-}1}$ is uncorrelated with $\kvec{w}_t$, which is used in prediction, and (ii) $\kvec{x}_{t|t{-}1}$  is uncorrelated with $\kvec{z}_t$, which is used in fusion, are easily proved to be correct by induction on $t$, using Lemma~\ref{lemma:mse}(ii). Figure~\ref{fig:dynamic-uncertainty} gives the intuition: $\kvec{x}_{t|t{-}1}$ for example is an affine function of the random variables $\kvec{x}_{0|0}, \kvec{w}_1, \kvec{v}_1, \kvec{w}_2, \kvec{v}_2,...,\kvec{w}_{t}$, and is therefore uncorrelated with $\kvec{v}_t$ (by assumption about $\kvec{v}_t$ and Lemma~\ref{lemma:mse}(ii)) and hence with $\kvec{z}_t$.

\begin{figure}[t]
	\centering
    % \hspace{-20pt}
	\includegraphics[scale=0.6]{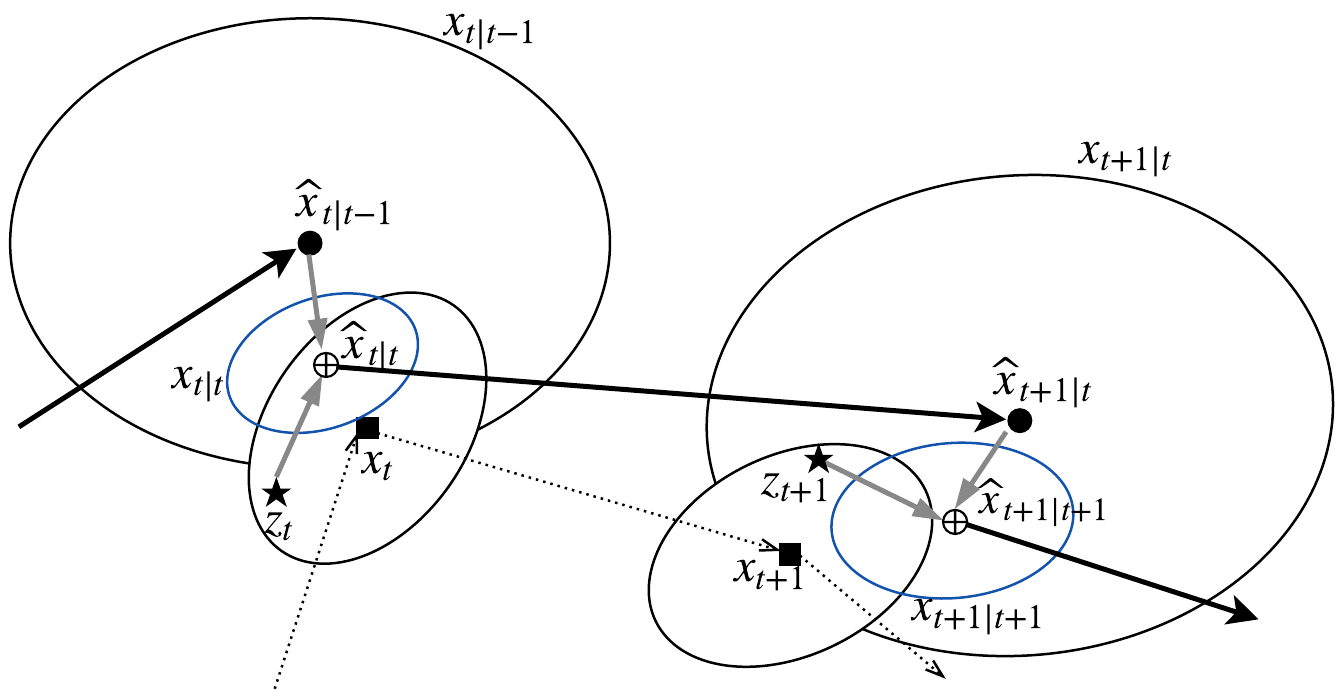}
    \caption{Pictorial representation of Kalman filtering.}
    \label{fig:kfstates}
\end{figure}

Figure~\ref{fig:kfstates} shows the computation pictorially using confidence ellipses to illustrate uncertainty. The dotted arrows at the bottom of the figure show the evolution of the state, and the solid arrows show the computation of the {\em a priori} estimates and their fusion with measurements.

%The state transition diagram of one time step shown in Figure~\ref{fig:kfstates}.
%The black arrows with solid lines represent the actual state flow.
%The steps in \kf are described by the dotted arrows and the gray arrows,
%representing the prediction and the fusion respectively.
%The light blue and dark blue eclipses indicate the uncertainty of the \emph{a
%priori} estimate and the \emph{a posteriori} estimate respectively, while the
%white eclipses show how noisy the measurements are.
%\francis{The description is partial}

\subsection{Fusing partial observations of the state}
\label{sec:Kalman-partial}

In some problems, only a portion of the state can be measured directly.
The observable portion of the state is specified formally using a full row-rank matrix $H_t$ called the {\em observation matrix}: if the state is $\kvec{x}$, what is observable is
$H_t\kvec{x}$. For example, if the state vector has two components and only the
first component is observable, $H_t$ can be $[\begin{smallmatrix} 1 & 0\end{smallmatrix}]$.
In general, the $H_t$ matrix can specify a linear relationship between
the state and the observation, and it can be time-dependent.
The imprecise measurement model introduced in Equation~\ref{eqn:observation_eqn} becomes:
\begin{align}
\label{eqn:observation_eqn_H}
  \kvec{z}_{t} = H_t\kvec{x}_{t} + \kvec{v}_{t}
\end{align}

The hidden portion of the state can be specified using
a matrix $C_t$, which is an orthogonal complement of $H_t$. For example, if
\mbox{$H_t= [\begin{smallmatrix} 1 & 0\end{smallmatrix}]$}, one choice for $C_t$
is \mbox{$[\begin{smallmatrix} 0 & 1\end{smallmatrix}]$}.

Figure~\ref{fig:dataflow-impl} shows the computation for this case. The fusion phase
can be understood intuitively in terms of the following steps.

\begin{closeenumeratei}
\item The observable part of the {\em a priori} estimate of the state
    $H_t\est{\kvec{x}}_{t|t-1}$ is fused with the measurement $\kvec{z}_{t}$,
    using the techniques developed in Sections~\ref{sec:scalar}-\ref{sec:vector}.
    The quantity \mbox{$\kvec{z}_{t} - H_t\est{\kvec{x}}_{t|t-1}$} is called the
    \emph{innovation}. The result is the {\em a posteriori} estimate of the observable state $H_t\est{\kvec{x}}_{t|t}$.

\item The BLUE in Section~\ref{sec:hidden} is used to obtain the
    \emph{a posteriori} estimate of the hidden state $C_t\est{\kvec{x}}_{t|t}$
    by adding to the \emph{a priori} estimate of the hidden state
    $C_t\est{\kvec{x}}_{t|t-1}$ a value obtained from the product of the covariance
    between $H_t\kvec{x}_{t|t-1}$ and $C_t\kvec{x}_{t|t-1}$ and the difference between
    $H_t\est{\kvec{x}}_{t|t-1}$ and $H_t\est{\kvec{x}}_{t|t}$.

    %from the \emph{a priori} estimate of the hidden state
    %($C_t\est{\kvec{x}}_{t|t-1}$), using the covariance between
    %$(H_t\kvec{x})$ and $(C_t\kvec{x})$ and the difference between
    %$(H_t\est{\kvec{x}}_{t|t-1})$ and $(H_t\est{\kvec{x}}_{t|t})$.

\item The \emph{a posteriori} estimates of the observable and hidden portions of
    the state are composed to produce the \emph{a posteriori} estimate of the
    entire state $\est{\kvec{x}}_{t|t}$.
\end{closeenumeratei}

The actual implementation produces the final result directly without going through these steps as shown in Figure~\ref{fig:dataflow-impl}, but these incremental steps are useful for understanding how all this works, and they are implemented as follows.

\begin{closeenumeratei}

\item The {\em a priori} estimate of the observable part of the state is $H_t\est{\kvec{x}}_{t|t-1}$
      and the covariance is $H_t\Sigma_{t|t-1} H_t^{\rm T}$.
      The {\em a posteriori} estimate is obtained directly from Equation~\ref{eqn:M1}:
\begin{align*}
  H_t\est{\kvec{x}}_{t|t} &= H_t\est{\kvec{x}}_{t|t-1} \\
                  &+ H_t\Sigma_{t|t-1} H_t^{\rm T}(H_t\Sigma_{t|t-1} H_t^{\rm T} + R_{t})^{-1} (\kvec{z}_{t} - H_t\est{\kvec{x}}_{t|t-1})
\end{align*}
\noindent
Let \mbox{$K_{t} {=} \Sigma_{t|t-1} H_t^{\rm T} (H_t \Sigma_{t|t-1} H_t^{\rm T} + R_{t})^{-1}$}. The {\em a posteriori} estimate
of the observable state can be written in terms of $K_{t}$ as follows:
\begin{align}
\label{eqn:h1}
H_t\est{\kvec{x}}_{t|t} &= H_t\est{\kvec{x}}_{t|t-1} + H_t K_{t} (\kvec{z}_{t} - H_t\est{\kvec{x}}_{t|t-1})
\end{align}

\item The {\em a priori} estimate of the hidden state is $C_t\est{\kvec{x}}_{t|t-1}$.
      The covariance between the hidden portion $C_t\kvec{x}_{t|t-1}$ and the observable
      portion $H_t\kvec{x}_{t|t-1}$ is $C_t\Sigma_{t|t-1} H_t^{\rm T}$.
      The difference between the \emph{a priori} estimate
      and \emph{a posteriori} estimate of $H_t\kvec{x}$ is $H_t K_{t} (\kvec{z}_{t} {-} H\est{\kvec{x}}_{t|t-1})$.
      Therefore the {\em a posteriori} estimate of the hidden portion of the state is obtained directly from Equation~\ref {eqn:blue1}:
\begin{align}
\nonumber
  C_t\est{\kvec{x}}_{t|t} &= C_t\est{\kvec{x}}_{t|t-1} \\
                          &+ (C_t \Sigma_{t|t-1} H_t^{\rm T})(H_t \Sigma_{t|t-1} H_t^{\rm T})^{-1}H_t K_{t}(\kvec{z}_{t} - H_t\est{\kvec{x}}_{t|t-1}) \nonumber \\
\label{eqn:c1}
&= C_t\est{\kvec{x}}_{t|t-1} + C_t K_{t}(\kvec{z}_{t} - H_t\est{\kvec{x}}_{t|t-1})
\end{align}

\item Putting the {\em a posteriori} estimates (\ref{eqn:h1}) and (\ref{eqn:c1}) together,
\begin{align*}
\begin{pmatrix} H_t \\ C_t \end{pmatrix} \est{\kvec{x}}_{t|t} &= \begin{pmatrix} H_t \\ C_t \end{pmatrix} \est{\kvec{x}}_{t|t-1} + \begin{pmatrix} H_t \\ C_t \end{pmatrix} K_{t} (\kvec{z}_{t} - H_t\est{\kvec{x}}_{t|t-1})
\end{align*}
\noindent
Since $\begin{pmatrix} H_t \\ C_t \end{pmatrix}$ is invertible, it can be canceled from the left and right hand sides, giving the equation
\begin{align}
\label{eqn:s1}
\est{\kvec{x}}_{t|t} &= \est{\kvec{x}}_{t|t-1} + K_{t}(\kvec{z}_{t} - H_t\est{\kvec{x}}_{t|t-1})
\end{align}
\end{closeenumeratei}

To compute $\Sigma_{t|t}$, Equation~\ref{eqn:s1} can be rewritten as\\
\mbox{$\est{\kvec{x}}_{t|t} = (I-K_{t}H_t)\est{\kvec{x}}_{t|t-1} + K_{t}\kvec{z}_{t}$}.
Since $\kvec{x}_{t|t-1}$ and $\kvec{z}_{t}$ are uncorrelated, it follows from Lemma~\ref{lemma:mse} that
\begin{align*}
\mbox{$\Sigma_{t|t} = (I-K_{t}H_t)\Sigma_{t|t-1}(I-K_{t}H_t)^{\rm T} + K_{t}R_{t}K_{t}^{\rm T}$}
\end{align*}

Substituting the value of $K_{t}$ and simplifying, we get
\begin{align}
\label{eqn:r1}
\Sigma_{t|t} = (I-K_{t}H_t)\Sigma_{t|t-1}
\end{align}

Figure~\ref{fig:dataflow-impl} puts all this together. In the literature, this dataflow is referred to as Kalman filtering. Unlike in Sections~\ref{sec:scalar} and \ref{sec:vector},
the Kalman gain is not a dimensionless value here. If $H_t = I$, the computations in Figure~\ref{fig:dataflow-impl} reduce to those of Figure~\ref{fig:dataflow-diag} as expected.

Equation~\ref{eqn:s1} shows that the {\em a posteriori} state estimate
is a linear combination of the {\em a priori} state estimate
($\est{\kvec{x}}_{t|t-1}$) and the measurement ($\kvec{z}_{t}$).
The optimality of this linear unbiased estimator is shown in the Appendix~\ref{sec:kfoptimality}.
In Section~\ref{sec:incr}, it was shown that incremental fusion of scalar estimates is optimal. The dataflow of Figures~\ref{fig:dynamics}(c,d) computes the {\em a posteriori} state estimate at time $t$ by incrementally fusing measurements from the previous time steps, and this incremental fusion can be shown to be optimal using a similar argument.

% \subsection{An example}
% \label{sec:example}
% \input{example}
\input{example}
\subsection{Discussion}
\label{sec:discussion}

This section shows that Kalman filtering for state estimation in linear systems
can be derived from two elementary ideas: optimal linear estimators for fusing uncorrelated estimates and best linear unbiased estimators for correlated variables. This is a different approach to the subject than the standard presentations in the literature. One standard approach is to use Bayesian inference. The other approach is to assume
that the {\em a posteriori} state estimator is a linear combination of the form $A_t \est{\kvec{x}}_{t|t-1} {+} B_t \kvec{z}_{t}$, and then find the values of $A_t$ and $B_t$ that produce an unbiased estimator with minimum \mse. We believe that the advantage of the presentation given here is that it exposes the concepts and assumptions that underlie
Kalman filtering.

Most presentations in the literature also begin by assuming that the
noise terms $\kvec{w}_t$ in the state evolution equation and
$\kvec{v}_t$ in the measurement are Gaussian. Some
presentations~\cite{kf-pictures,faragher2012} use properties of Gaussians to
derive the results in Sections~\ref{sec:scalar} although as we have
seen, these results do not depend on distributions being Gaussians.
Gaussians however enter the picture in a deeper way if one considers
\emph{nonlinear} estimators. It can be shown that if the noise terms
are not Gaussian, there may be nonlinear estimators whose \mse is
lower than that of the linear estimator presented in
Figure~\ref{fig:dataflow-impl}. However if the noise is Gaussian, this
linear estimator is as good as any unbiased nonlinear estimator (that
is, the linear estimator is a \emph{minimum variance unbiased
estimator} (MVUE)). This result is proved using the Cramer-Rao lower
bound~\cite{rao45}.

%% file: example.tex
\subsection{Example: falling body}
\label{sec:example}

\begin{figure*}[t]
	\centering
	\subfloat[Evolution of state: Distance]{\includegraphics[scale=0.6]{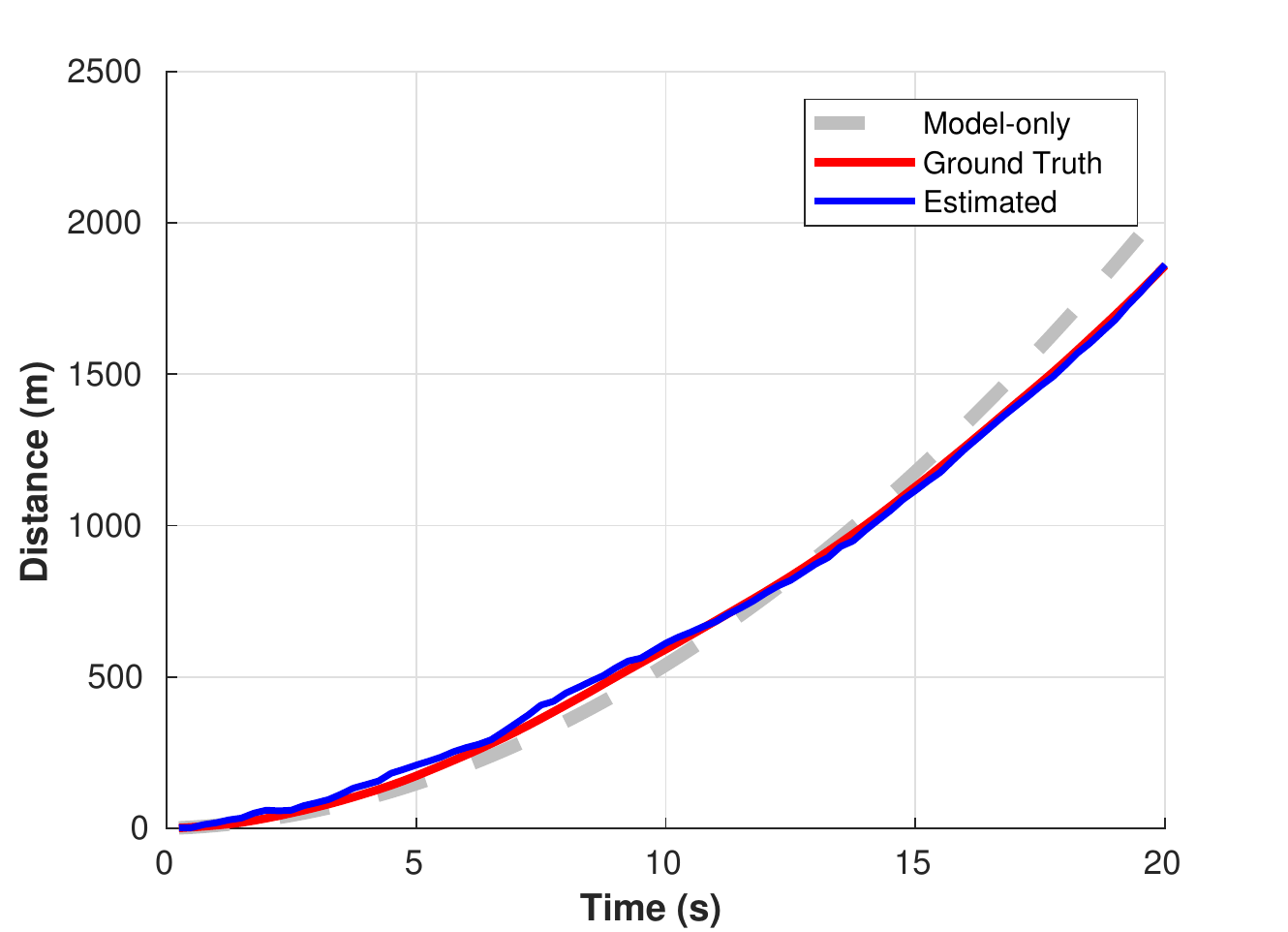}\label{fig:d_plot}}
	\hspace*{32pt}
	\subfloat[Evolution of state: Velocity]{\includegraphics[scale=0.6]{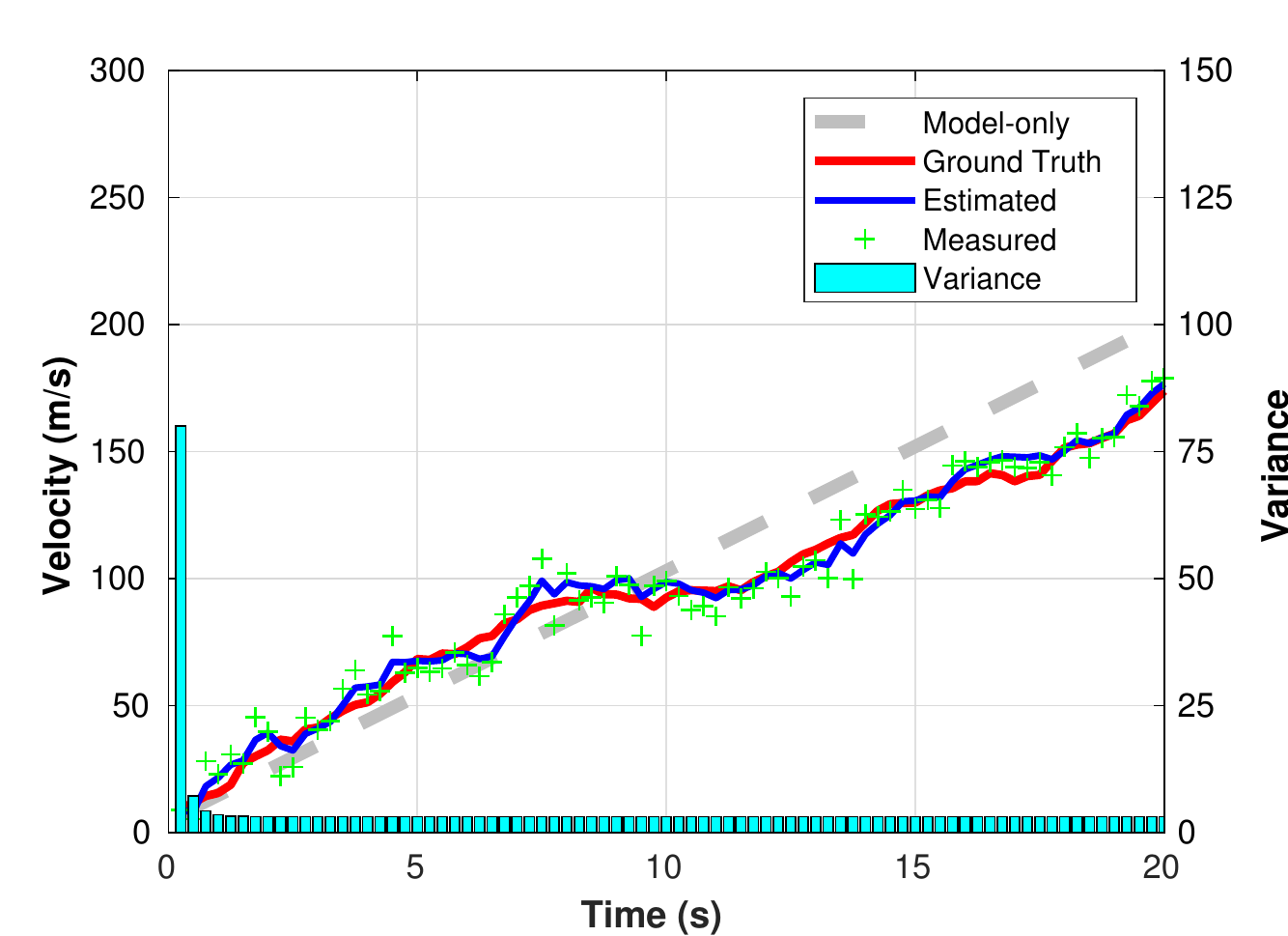}\label{fig:v_plot}}
    \caption{Estimates of the object's state over time.}
    \label{fig:car_motion_result}
\end{figure*}

To demonstrate the effectiveness
of the \kf, we consider an example in which an object falls
from the origin at time $t{=}0$ with an initial speed of $0$ m/s and an expected
constant acceleration of $9.8$ m/s$^{2}$ due to gravity. Note that acceleration
in reality may not be constant due to factors such as wind, air friction, and so
on.

The state vector of the object contains two components, one for the distance
from the origin $s(t)$ and one for the velocity $v(t)$. We assume that only the
velocity state can be measured at each time step. If time is discretized in
steps of 0.25 seconds, the difference equation for the dynamics of the system is easily shown to be the following:
\begin{align}
\label{eqn:car}
\begin{pmatrix} v_{n} \\ s_{n} \end{pmatrix} &= \begin{pmatrix}1 & 0 \\ 0.25 & 1\end{pmatrix}\begin{pmatrix} v_{n-1} \\ s_{n-1} \end{pmatrix}  + \begin{pmatrix} 0 & 0.25 \\ 0 & 0.5\times0.25^2\end{pmatrix} \begin{pmatrix} 0 \\ 9.8 \end{pmatrix}
\end{align}
%where \mbox{$\begin{pmatrix} v_0 \\ s_0\end{pmatrix} = \begin{pmatrix} 0 \\ 0\end{pmatrix}$}.
where we assume \mbox{$\begin{pmatrix} v_0 \\ s_0\end{pmatrix} = \begin{pmatrix} 0 \\ 0\end{pmatrix}$} and $\Sigma_0={\begin{pmatrix}80 & 0 \\ 0 & 10\end{pmatrix}}$.

The gray lines in Figure~\ref{fig:car_motion_result} show the evolution of velocity and distance with time according to this model. Because of uncertainty in modeling the system dynamics, the actual evolution of the velocity and position will be different in practice. The red lines in Figure~\ref{fig:car_motion_result} show one trajectory for this evolution, corresponding to a Gaussian noise term
with covariance $\begin{pmatrix}2 & 2.5 \\ 2.5 & 4\end{pmatrix}$ in Equation~\ref{eqn:state_evolution} (because this noise term is random, there are many trajectories for the evolution, and we are just showing one of them). The red lines correspond to ``ground truth'' in our example.

The green points in Figure~\ref{fig:v_plot} show the noisy measurements of velocity at different time steps, assuming the noise is modeled by a Gaussian with variance $8$. The blue lines show the {\em a posteriori} estimates of the velocity and position. It can be seen that the {\em a posteriori} estimates track the ground truth quite well even when the ideal system model (the gray lines) is inaccurate and the measurements are noisy. The cyan bars in the right figure show the variance of the velocity at different time steps. Although the initial variance is quite large, application of Kalman filtering is able to reduce it rapidly in few time steps.

%% file: extension.tex
The \emph{Extended Kalman Filter} (EKF) and \emph{Unscented Kalman Filter} (UKF)
are heuristic approaches to using Kalman filtering for nonlinear systems.
The state evolution and measurement equations for nonlinear systems
with additive noise can be written as follows; in these equations,
$f$ and $h$ are nonlinear functions.
\begin{align}
\label{eqn:nl1}
\kvec{x}_{t} &= f(\kvec{x}_{t-1}, \kvec{u}_t) + \kvec{w}_t \\
\label{eqn:nl2}
\kvec{z}_{t} &= h(\kvec{x}_{t}) + \kvec{v}_{t}
\end{align}

Intuitively, the EKF  constructs linear approximations to the nonlinear
functions $f$ and $h$ and applies the Kalman filter equations,
while the UKF constructs approximations to probability distributions
and propagates these through the nonlinear functions to construct approximations
to the posterior distributions.

\paragraph{EKF}
Examining Figure~\ref{fig:dataflow-impl}, we see that the {\em a priori} state
estimate in the predictor can be computed using the system model:
\mbox{$\est{\kvec{x}}_{t|t-1} = f(\est{\kvec{x}}_{t-1|t-1}, \kvec{u}_t)$}.
However, since the system dynamics and measurement equations are nonlinear,
it is not clear how to compute the covariance matrices for the {\em a priori}
estimate and the measurement. In the EKF, these
matrices are computed by linearizing Equations~\ref{eqn:nl1} and \ref{eqn:nl2}
using the Taylor series expansions for the nonlinear
functions $f$ and $h$. This requires computing
the following \emph{Jacobians}\footnote{The Jacobian matrix is
the matrix of all first order partial derivatives of a vector-valued function.},
which play the role of $F_t$ and $H_t$ in Figure~\ref{fig:dataflow-impl}.

\begin{align*}
F_t = \frac{\partial f}{\partial \kvec{x}}\bigg\rvert_{\est{\kvec{x}}_{t-1|t-1},\kvec{u}_t}
H_t = \frac{\partial h}{\partial \kvec{x}}\bigg\rvert_{\est{\kvec{x}}_{t|t-1}}
\end{align*}

The EKF performs well in some applications such as navigation systems
and GPS~\cite{Thrun:2005}.

%Though the state estimation can be performed using $f_t$ without any problem, the uncertainty
%distribution will get distorted if the propagation transformation is not linear.
%Hence, as discussed in Section~\ref{sec:discussion}, \kf with Gaussian noise
%distribution will lose its optimality compared to non-linear estimators
%after uncertainty propagation if $F_t$ or $H$ is non linear.
%of $f_t$ and $h_t$ with respect to the state value, $F'_t$ and $H'_t$, are calculated at each time step for
%uncertainty update. The EKF algorithm is very similar to the regular \kf in
%Figure~\ref{fig:dataflow-impl}, except that $f_t$ and $h_t$ need to be
%linearized at each time step.
%In the case of well defined transition models,
%Note that the noise will remain Gaussian if the transformation is linear.
%\footnote{\url{https://en.wikipedia.org/wiki/Extended\_Kalman\_filter}}

\paragraph{UKF}
When the system dynamics and observation models are highly nonlinear, the
Unscented Kalman Filter (UKF)~\cite{Julier04} can be an improvement over the EKF. The UKF
is based on the {\em unscented transformation}, which is a method for computing
the statistics of a random variable $\kvec{x}$ that undergoes a nonlinear transformation
($\kvec{y}=g(\kvec{x})$). The random variable $\kvec{x}$ is sampled using a carefully chosen
set of {\em sigma points} and these sample points are propagated through the
nonlinear function $g$. The statistics of $\kvec{y}$ are estimated using a weighted
sample mean and covariance of the posterior sigma points. The UKF tends to be more
robust and accurate than the EKF but has higher computation overhead due to the
sampling process.

%\footnote{\url{https://en.wikipedia.org/wiki/Kalman\_filter\#Unscented\_Kalman\_filter}}

%%% Local Variables:
%%% mode: latex
%%% TeX-master: "paper"
%%% End:

%% file: conclusions.tex
In this paper, we have shown that two concepts - optimal linear estimators for fusing uncorrelated estimates and best linear unbiased estimators for correlated variables -
provide the underpinnings for Kalman filtering. By combining these ideas, standard results on Kalman filtering for linear systems can be derived in an intuitive and straightforward way that is simpler than other presentations of this material in the literature. This approach makes clear the assumptions that underlie the optimality results associated with Kalman filtering, and should make it easier to apply Kalman filtering to problems in computer systems.

%%% Local Variables:
%%% mode: latex
%%% TeX-master: "paper"
%%% End:

%% file: appendix.tex
\section{Basic probability theory and statistics terminology}{}
\label{sec:basic_terms}

\paragraph{Probability density function} For a continuous random variable $x$, a \emph{probability density function} (pdf) is a function $p(x)$ whose value provides a relative likelihood that the value of the random variable will equal $x$. The integral of the pdf within a range of values is the probability that the random variable will take a value within that range.

If $g(x)$ is a function of $x$ with pdf $p(x)$, the {\em expected value} or {\em expectation} of $g(x)$ is $E[g(x)]$, defined as the following integral:
\begin{align*}
E[g(x)] = \int_{-\infty}^{\infty} g(x) p(x)dx
\end{align*}

By definition, the \emph{mean} $\mu_x$ of a random variable $x$ is $E[x]$. The \emph{variance} of a random variable x measures the variability of the distribution. For the set of possible values of $x$, variance (denoted by $\sigma^2_x$) is defined by $\sigma^2_x = E[(x-\mu_x)^2]$. The variance of a continuous random variable $x$ can be written as the following integral:
\begin{align*}
\sigma^2_x = \int_{-\infty}^{\infty} (x-\mu)^2 p(x)dx
\end{align*}

If $x$ is discrete and all outcomes are equally likely, then $ \sigma^2_x = \frac{\Sigma (x_i - \mu_x)^2}{n}$. The standard deviation $\sigma_x$ is the square root of the variance.\

\paragraph{Covariance} The \emph{covariance} of two random variables is a measure of their joint variability. The covariance between random variables $x_1:p_1{\sim}(\mu_1,\sigma_1^2)$ and $x_2:p_2{\sim}(\mu_2,\sigma_2^2)$ is the expectation $E[(x_1{-}\mu_1){*}(x_2{-}\mu_2)]$. Two random variables are \emph{uncorrelated} or \emph{not correlated} if their covariance is zero. This is not the same concept as {\em independence} of random variables.

Two random variables are independent if knowing the value of one of the variables does not
give us any information about the possible values of the other one. This is written formally as \mbox{$p(x_1|x_2) = p(x_1)$}; intuitively, knowing the value of $x_2$ does not change the probability that $p_1$ takes a particular value.

Independent random variables are uncorrelated but random variables can be uncorrelated even if they are not independent. It can be shown that if $x_1$ and $x_2$ are not correlated, \mbox{$E[x_1|x_2] {=} E[x_1]$}; intuitively, knowing the value of $x_2$ may change the probability that $x_1$ takes a particular value, but the mean of the resulting distribution
remains the same as the mean of $x_1$. A special case of this that is easy to understand are examples in which knowing $x_2$ restricts the possible values of $x_1$ without changing the mean. Consider a random variable $u:U$ that is uniformly distributed over the unit circle, and consider random variables $x_1:[-1,1]$ and $x_2:[-1,1]$ that are the projections of $u$ on the $x$ and $y$ axes respectively. Given a value for $x_2$, there are only two possible values for $x_1$, so $x_1$ and $x_2$ are not independent. However, the mean of these values is 0, which is the mean of $x_1$, so $x_1$ and $x_2$ are not correlated.

\section{Matrix Derivatives}
\label{sec:matrix_derivative}
If $f(\kmat{X})$ is a scalar function of a matrix $\kmat{X}$, the matrix derivative
$\frac{\partial f(\kmat{X})}{\partial \kmat{X}}$ is defined as the matrix

\begin{center}
$\left(
  \begin{array}{ccc}
    \frac{\partial f(\kmat{X})}{\partial X(1,1)}  & ... & \frac{\partial f(\kmat{X})}{\partial X(1,n)} \\
    ... & ... & ... \\
    \frac{\partial f(\kmat{X})}{\partial X(n,1)} & ... & \frac{\partial f(\kmat{X})}{\partial X(n,n)}
  \end{array}
\right)$
\end{center}

\begin{lemma}
Let \kmat{X} be a $m\times n$ matrix, \kvec{a} be a $m\times1$ vector, \kvec{b} be a $n\times1$ vector.

\begin{align}
\label{eqn:vecDerivative}
\frac{\partial \kvec{a}^T \kmat{X} \kvec{b}}{\partial \kmat{X}} &= \kvec{a} \kvec{b}^T \\
\label{eqn:matDerivative}
\frac{\partial (\kvec{a}^T \kmat{X}^T \kmat{X} \kvec{b})}{\partial \kmat{X}} &= \kmat{X}(\kvec{a}\kvec{b}^T + \kvec{b} \kvec{a}^T) \\
\label{eqn:trace2}
\frac{\partial (trace(XBX^{T}))}{\partial X} &= XB^{T} + XB
\end{align}

%\begin{proof}
%We sketch the proofs of both parts below.
%
%\begin{itemize}
%\item Equation~\ref{eqn:vecDerivative}: In this case, $f(\kmat{X}) = \kvec{a}^{\rm T} \kmat{X} \kvec{b}$.\\
%\mbox{$\frac{\partial f(\kmat{X})}{\partial X(i,j)} = \kvec{a}(i)\kvec{b}(j) = (\kvec{a}\kvec{b}^T)(i,j)$.}
%
%\item Equation~\ref{eqn:matDerivative}: In this case, $f(\kmat{X}) = \kvec{a}^{\rm T} \kmat{X}^{\rm T} \kmat{X} \kvec{b} = (\kmat{X}\kvec{a})^{\rm T} \kmat{X} \kvec{b}$, which is equal to
%\begin{align*}
%\smashoperator[r]{\sum_{i=1}^{m}} \big(\smashoperator[r]{\sum_{k=1}^{n}}X(i,k){*}a(k)\big) * \big(\smashoperator[r]{\sum_{k=1}^{n}}X(i,k){*}b(k)\big)
%\end{align*}
%
%\noindent
%Therefore,
%\begin{align*}
%    \frac{\partial f(\kmat{X})}{\partial X(i,j)} &= a(j)*\smashoperator[r]{\sum_{k=1}^{n}}X(i,k){*}b(k) \\
%                                                 &+ b(j)*\smashoperator[r]{\sum_{k=1}^{n}}X(i,k){*}a(k)
%\end{align*}
%
%It is easy to see that this is the same value as the $(i,j)^{th}$ element of \mbox{$\kmat{X}(\kvec{a}\kvec{b}^T + \kvec{b} \kvec{a}^T)$}.
%\end{itemize}
%\end{proof}

\end{lemma}

See Petersen and Pedersen for a proof~\cite{matrix-cookbook}.

\section{Proof of Theorem~\ref{th:multipleVec}}
\label{sec:proofMultipleVec}

Theorem~\ref{th:multipleVec}, which is reproduced below for convenience,
can be proved using matrix derivatives.

\begin{prop*}
Let pairwise uncorrelated estimates $\kvec{x}_i (1 {\leq} i {\leq} n)$ drawn from distributions
$p_i(x){=}(\pmb{\mu}_i,{\Sigma}_i)$ be fused using the linear model
$\kvec{y}_A(\kvec{x}_1,..,\kvec{x}_n) = \sum_{i=1}^n \kmat{A}_i \kvec{x}_i$, where $\sum_{i=1}^n \kmat{A}_i = \kmat{I}$. The $MSE(\kvec{y}_A)$ is minimized for
\begin{align*}
\kmat{A}_i &=  (\smashoperator[r]{\sum_{{\rm j}=1}^n} {\Sigma}_j^{-1})^{-1} {\Sigma}_i^{-1}.
\end{align*}
\end{prop*}

\begin{proof}

To use the Lagrange multiplier approach,
we can convert the constraint $\sum_{i=1}^n \kmat{A}_i = \kmat{I}$ into a set of
$m^2$ scalar equations (for example, the first equation would be $A_1(1,1) + A_2(1,1)+..+A_n(1,1) = 1$), and then introduce $m^2$ Lagrange multipliers, which can denoted by $\lambda(1,1),...\lambda(m,m)$.

This obscures the matrix structure of the problem so it is better to implement this
idea implicitly. Let ${\Lambda}$ be an $m{\times}m$ matrix in which each entry is one of the scalar Lagrange multipliers we would have introduced in the approach described above. Analogous to the inner product of vectors, we can define the inner product of two
matrices as ${<}A,B{>} = trace(A^{\rm T}B)$ (it is easy to see that ${<}A,B{>}$ is $\sum_{i=1}^m\sum_{j=1}^m A(i,j)B(i,j)$). Using this notation, we can formulate the optimization problem using Lagrange multipliers as follows:
\begin{align*}
f(\kmat{A}_1,...,\kmat{A}_n) = E\big\{ \smashoperator[r]{\sum_{i=1}^n}
(\kvec{x}_i-\pmb{\mu}_i)^{\rm T}&{\kmat{A}_i}^{\rm T} \kmat{A}_i (\kvec{x}_i-\pmb{\mu}_i)\big\} \\
                             &+ \big<{\Lambda},\big(\smashoperator[r]{\sum_{i=1}^n} \kmat{A}_i - {I}\big)\big>
\end{align*}

Taking the matrix derivative of $f$ with respect to each \kmat{A$_i$} and setting each derivative to zero to find the optimal values of $A_i$ gives us the equation
\mbox{$E\big\{ 2\kmat{A}_i(\kvec{x}_i-\pmb{\mu}_i)(\kvec{x}_i-\pmb{\mu}_i)^{\rm T} + {\Lambda}\big\} = 0$}.

This equation can be written as
\mbox{$2\kmat{A}_i {\Sigma}_i + {\Lambda} = 0$}, which implies
\begin{align*}
\kmat{A}_1{\Sigma}_1 = \kmat{A}_2{\Sigma}_2 = ... = \kmat{A}_n{\Sigma}_n = -\frac{{\Lambda}}{2}
\end{align*}

Using the constraint that the sum of all \kmat{A$_i$} equals to the identity matrix ${I}$ gives us the desired expression for $\kmat{A}_i$:
\begin{align*}
\kmat{A}_i &= (\smashoperator[r]{\sum_{{\rm j}=1}^n} {\Sigma}_j^{-1})^{-1} {\Sigma}_i^{-1}
\end{align*}
\end{proof}

%\section*{Proof of the Optimality of Equation~\ref{eqn:s0}}
%\label{sec:hcoptimality}
%
%\begin{align}
%MSE(\begin{pmatrix} H_t \\ C_t \end{pmatrix} \kvec{x}_{t|t}) &= E[(\kvec{x}_{t|t} - \est{\kvec{x}}_{t|t})^T \begin{pmatrix} H_t {~} C_t \end{pmatrix}^T \begin{pmatrix} H_t \\ C_t \end{pmatrix} (\kvec{x}_{t|t} - \est{\kvec{x}}_{t|t})] \nonumber
%\\
%&= E[(\kvec{x}_{t|t} - \est{\kvec{x}}_{t|t})^T (H_t^T H_t + C_t^T C_t) (\kvec{x}_{t|t} - \est{\kvec{x}}_{t|t})] \nonumber
%\\
%\label{eqn:hc_decomp}
%&= MSE(H_t\kvec{x}_{t|t}) + MSE(C_t\kvec{x}_{t|t})
%\end{align}
%
%The MSE of $\begin{pmatrix} H_t \\ C_t \end{pmatrix}\kvec{x}_{t|t}$ can be decomposed as
%the sumation of the MSE of $H_t\kvec{x}_{t|t}$ and $C_t\kvec{x}_{t|t}$, as shown
%in Equation~\ref{eqn:hc_decomp}. Since
%$H_t K_t$ and $C_t K_t$ minimize the MSE of $H_t\kvec{x}_{t|t}$ and $C_t\kvec{x}_{t|t}$
%respectively, $\begin{pmatrix} H_t \\ C_t \end{pmatrix}K_t$ minimizes the MSE of
%$\begin{pmatrix} H_t \\ C_t \end{pmatrix}\kvec{x}_{t|t}$.

\section{Proof of the Optimality of Equation~\ref{eqn:s1}}
\label{sec:kfoptimality}

We show that \mbox{($\est{\kvec{x}}_{t|t} = \est{\kvec{x}}_{t|t-1} + K_{t}(\kvec{z}_{t} - H_t\est{\kvec{x}}_{t|t-1})$)} (Equation~\ref{eqn:s1}) is an optimal unbiased linear estimator
for fusing the {\em a priori} state estimate with the measurement at each step.
The proof has two steps: we show that this estimator is unbiased, and then show it is optimal.

\paragraph{Unbiased condition:}
We prove a more general result that characterizes unbiased linear estimators for this problem, assuming that the prediction stage (Figure~\ref{fig:dynamics}(d)) is unchanged. The general form of the linear estimator for computing the {\em a posteriori} state estimate is
\begin{align}
\label{eqn:kfGeneral}
\est{\kvec{y}}_{t|t} = A_{t}*\est{\kvec{y}}_{t|t-1} + B_{t}*\kvec{z}_{t}
\end{align}
It is unbiased if \mbox{$E[\est{\kvec{y}}_{t|t}] {=} E[\kvec{x}_t]$}, and we show that this is true if \mbox{$A_{t} {=} {(}I{-}B_{t}{)}*H_t$}.

The proof is by induction on $t$. By assumption, \mbox{$E[\est{\kvec{y}}_{0|0}] = E[\kvec{x}_0]$}. Assume inductively that \mbox{$E[\est{\kvec{y}}_{t{-}1|t{-}1}] {=} E[\kvec{x}_{t{-}1}]$}.

\noindent
(a) We first prove that the predictor is unbiased.
\begin{align*}
\est{\kvec{y}}_{t|t{-}1} &= \kmat{F}_t{*}\est{\kvec{y}}_{t{-}1|t{-}1} + \kmat{B}_t{*}\kvec{u}_t \text{\ \ (Predictor in Figure~\ref{fig:dynamics})}\\
E[\est{\kvec{y}}_{t|t-1}] &= \kmat{F}_t*E[\est{\kvec{y}}_{t{-}1|t{-}1}] + \kmat{B}_t{*}\kvec{u}_t \\
&= \kmat{F}_t*E[\kvec{x}_{t-1}] + \kmat{B}_t{*}\kvec{u}_t \text{\ \ (By inductive assumption)} \\
&= E[\kmat{F}_t*\kvec{x}_{t-1} + \kmat{B}_t{*}\kvec{u}_t] \\
&= E[\kmat{F}_t*\kvec{x}_{t-1} + \kmat{B}_t{*}\kvec{u}_t + \kvec{w}_t] \ \ \text{($\kvec{w}_t$ is zero mean)}\\
&= E[\kvec{x}_t] \text{\ \ (From state evolution equation~\ref{eqn:state_evolution})}
\end{align*}

\noindent
(b) We prove that the estimator in Equation~\ref{eqn:kfGeneral} is unbiased if $A_{t}{=} {(}I{-}B_{t}{)}*H_t$.
\begin{align*}
E[\est{\kvec{y}}_{t|t}] &= E[A_{t}*\est{\kvec{y}}_{t|t-1} + B_{t}*\kvec{z}_{t}]\ \  \text{(From Equation~\ref{eqn:kfGeneral})}\\
&= A_{t}*E[\est{\kvec{y}}_{t|t-1}] + B_{t}*E[\kvec{z}_{t}] \\
&= A_{t}*E[\kvec{x}_t] + B_{t}*E[H_t\kvec{x}_{t} + \kvec{v}_{t}] \ \ \text{(Equation~\ref{eqn:observation_eqn_H} for $\kvec{z}_t$)} \\
&= A_{t}*E[\kvec{x}_t] + B_{t}*H_t*E[\kvec{x}_t]\ \  \text{(Because $\kvec{v}_t$ is zero mean)} \\
&= (A_{t} + B_{t}*H_t)*E[\kvec{x}_t]
\end{align*}
The estimator is unbiased if \mbox{$(A_{t} + B_{t}*H_t) = I$}, which is equivalent to requiring that \mbox{$A_{t} = (I-B_{t}*H_t)$}. Therefore the general unbiased linear estimator is of the form
\begin{align}
\est{\kvec{y}}_{t|t} &= (I-B_{t}*H_t)*\est{\kvec{y}}_{t|t-1} + B_{t}*\kvec{z}_{t} \nonumber \\
\label{eqn:kfUnbiased}
&= \est{\kvec{y}}_{t|t-1} + B_{t}*(\kvec{z}_{t} - H_t*\est{\kvec{y}}_{t|t-1})
\end{align}

Since Equation~\ref{eqn:s1} is of this form, it is an unbiased linear estimator.

\paragraph{Optimality:} We now show that using $B_t = K_t$ at each step is optimal, assuming that this is done at all time steps before $t$.
Since $\kvec{z}_{t}$ and $\kvec{y}_{t|t-1}$ are uncorrelated, we can use Lemma~\ref{lemma:mse} to compute the covariance matrix of $\est{\kvec{y}}_{t|t}$, denoted by
$\Xi_{t|t}$. This gives
\mbox{$\Xi_{t|t} = (I-B_{t}H_t)\Sigma_{t|t-1}(I-B_{t}H_t)^{\rm T} + B_{t}R_{t}B_{t}^{\rm T}$}.
The $MSE$ is the trace of this matrix, and we need to find $B_t$ that minimizes this trace. Using matrix derivatives (Equation~\ref{eqn:trace2}), we see that
\begin{align*}
\frac{\partial(trace(\Xi_{t|t}))}{\partial B_t} = -2(I-B_tH_t)\Sigma_{t|t-1}H_t^{T} + 2*B_tR_t
\end{align*}

Setting this zero and solving for $B_t$ gives \newline \mbox{$B_t = \Sigma_{t|t-1}H_t^{T}[H_t\Sigma_{t|t-1}H_t^{T}+R_t]^{-1}$}. This is exactly $K_t$, proving that Equation~\ref{eqn:s1} is an optimal unbiased linear estimator.

\paragraph{Comment:} This proof of optimality provides another way of deriving Equation~\ref{eqn:s1}. We believe the constructive approach described in Section~\ref{sec:dynamics} provides more insight into how and why Kalman filtering works. The ideas in that construction can be used to provide a different proof of optimality.